\documentclass[12pt]{article}
\usepackage[utf8]{inputenc}
\usepackage{hyperref}
\usepackage{geometry}
\usepackage{graphicx}
\usepackage{booktabs}
\usepackage{longtable}
\usepackage{caption}
\usepackage{float}

\title{Studying Disinformation Narratives on Social Media with LLMs and Semantic Similarity}
\author{Chaytan Chief Inman\\
Master of Arts in International Studies\\
University of Washington\\
Committee: Jessica L. Beyer, Tadayoshi Kohno}

\date{2025}

\begin{document}

\maketitle

\begin{abstract}
This thesis develops a continuous scale measurement of similarity to disinformation narratives that can serve to detect disinformation and capture the nuanced, partial truths that are characteristic of it. To do so, two tools are developed and their methodologies are documented. The tracing tool takes tweets and a target narrative, rates the similarities of each to the target narrative, and graphs it as a timeline. The second narrative synthesis tool clusters tweets above a similarity threshold and generates the dominant narratives within each cluster. These tools are combined into a Tweet Narrative Analysis Dashboard. The tracing tool is validated on the GLUE STS-B benchmark, and then the two tools are used to analyze two case studies for further empirical validation. The first case study uses the target narrative “The 2020 election was stolen” and analyzes a dataset of Donald Trump’s tweets during 2020. The second case study uses the target narrative, “Transgender people are harmful to society” and analyzes tens of thousands of tweets from the media outlets The New York Times, The Guardian, The Gateway Pundit, and Fox News. Together, the empirical findings from these case studies demonstrate semantic similarity for nuanced disinformation detection, tracing, and characterization.

The tools developed in this thesis are hosted at \url{https://narrativedashboard.xyz} and can be accessed through the permission of the author. Please explain your use case in your request. The HTML friendly version of this paper is at \url{https://chaytanc.github.io/projects/disinfo-research/} (Inman, 2025).

\end{abstract}

\section*{Acknowledgements}
To the following people, I am so grateful for your help guiding this thesis and my path leading up to it: 

\begin{center}
    Jessica Beyer, Yoshi Kohno, Stephen Prochaska. \\
    Thank you! \textless3
\end{center}

\tableofcontents

\section{Introduction}
\subsection{Problem}

{Disinformation}{~-- or information that is intentionally misleading or
false and with intent to harm (Wardle \& Derakhshan, 2017) --
proliferates on social }{media}{~(Subcommittee on Intelligence and
Special Operations, 2021) but mingles among millions of authentic
opinions. The effects are difficult to measure (Rid, 2021) but one clear
example is the effect of Donald Trump's tweets about the US 2020
presidential election, which undermined public trust in democracy and
culminated in riots at the Capitol Building on January 6th 2021 (Bowler,
Carreras \& Merolla, 2022; Fried \& Harris 2020). Given these stakes, it
is important t}{o}{~understand the problem -- to characterize
disinformation, track its spread, attribute its source, identify its
causes, and assess its impacts -- which inevitably requires an initial
classification or form of detection. But, as Thomas Rid (2021) argues,
the nature of disinformation means it mixes lies with the truth in
subtle ways (Rid, 2021). }{Rid examines the slippery nature of
disinformation and describes it in the following way:}

{Disinformation works, and in unexpected ways. The fine line between
fact and forgery may be clear in the moment an operator or an
intelligence agency commits the act of falsification\ldots{} or when a
bogus online account invites unwitting users to join a street
demonstration, or shares extremist posts. But fronts, forgeries and
fakes don't stop there\ldots{} When social media users gather in the
streets following a bogus event invitation, the demonstration is
real\ldots{} Engineered effects were very difficult to isolate from
organic developments.}

{}

{Thus, detecting disinformation requires a nuanced understanding of
context, actors, and intentions, even in a world of big data.}

{}

{Detecting and characterizing disinformation on social media has
methodological blindspots that tend to overlook nuance in favor of large
scale data analysis. Current methods for detecting disinformation for
analysis tend to fall into two categories (Kennedy et al., 2022). The
first method is using keyword searches to map data to disinformation
narratives, which can handle large amounts of data that social media
produces, but fails to understand nuance. The second is relying on
researchers to perform qualitative coding and evaluation, which limits
the focus to smaller amounts of data and is very time intensive. Mixed
methods also exist, such as the extremely thorough process developed by
Kennedy et al. (2022) which involved 112 researchers across multiple
organizations to hand-curate possible disinformation narratives. }

{}

{Yet all of these methods tend to grapple with trading a deep
understanding of how texts function as disinformation for simplified big
data analysis or vice versa. I'll examine the particularly thorough
approach of Kennedy et al. (2022) to see this tradeoff in action. In
their mixed methods approach, the method of Kennedy et al. employed
their hand-curated disinformation narratives to tease out differences in
the character of disinformation spread (as opposed to a binary
classification of disinformation or not). They found 456 distinct
disinformation narratives. However, they still relied on keyword
searches to classify tweets within the 456 disinformation narratives,
since their dataset contained over 30 million tweets. Because such
keyword searches contain misclassifications, they applied a process of
random sampling and hand-evaluating the ``noise'' (misclassifications)
in the dataset. While Kennedy et al. did not simply classify tweets as
disinformation or not, it was also a process that required a 112 person
team of researchers to achieve. Moreover, }{visualizing the broad spread
of disinformation over time becomes difficult with 456}{~disinformation
narrative categories. When presenting their results, the nuance was
necessarily distilled into a graph of ``Misinformation Tweets per Day''
for one particular disinformation narrative. If one wants to narrow the
scope of analysis in order to maintain human supervision of this
classification process or simplify visualization, the case study in this
paper shows that even a singular user's tweets (Donald Trump's) number
in the hundreds of thousands. It is therefore easy to see how a nuanced
analysis can be sacrificed by the amount of data necessary to analyze
and the desire to visualize broad trends}{.}{~}

{}

{Natural language processing, and particularly the process of embedding
meaning from sentences and longer contexts into a numerical space,
offers a potential method of detecting disinformation at scale while
still characterizing the ambiguous, partially truthful nature of text
related to disinformation. One way of doing this is by using the
continuous scale measurement of semantic similarity between known
disinformation narratives and potential disinformation narratives.
Semantic similarity describes metrics that estimate similarity of the
meaning of texts (Chandrasekaran \& Mago, 2021). In the context of LLMs
and in this paper, this often refers to a cosine similarity of
embeddings ~(Chandrasekaran \& Mago, 2021). To explore this possibility,
t}{he}{~question this research tackles is, ``Can we use a semantic
similarity metric derived from large language models (LLMs) to
quantitatively trace and characterize disinformation narratives on
social media on a continuous, non-binary scale?'' }

\subsection{Paper Outline}

{To answer this question, I begin by proposing a framework of five
foundational challenges in the field of disinformation, and then use
this framework to review the existing methodologies in disinformation
}{research}{. In the Background, I overview the fundamentals of large
language models as a basis for the disinformation detection tools and
methodology developed in this paper. I also examine current research on
quantitative analysis of social media and disinformation detection to
clarify the research contributions made in this paper. In the Methods
section, I develop a ``tracing tool'' and a ``narrative synthesis tool''
using a large language model derived metric of semantic similarity.
Then, I combine }{the two tools }{into a Tweet Narrative Analysis
Dashboard to trace and characterize disinformation narratives. In order
to give confidence that the semantic similarity metric is meaningful, I
evaluate the model underlying the tracing tool on the popular NLP
reasoning benchmark, the General Language Understanding Evaluation
(GLUE) (Wang et al., 2019) Semantic Textual Similarity Benchmark (STS-B)
(Cer et al., 2017). The STS-B measures the extent to which semantic
similarity is able to match human judgement of textual similarity. The
GLUE dataset is used for method validation on a standard NLP benchmark
and the examples in it are not related to the empirical results from the
following case studies. Finally, I demonstrated these tools }{in}{~two
case studies of disinformation on social media -- tracing the 2020
election }{hoax }{narrative in Donald Trump's tweets, and characterizing
and tracing the narrative that transgender people are harmful to
society. Therefore, in answering this research question, this paper
makes a methodological contribution to disinformation research and
demonstrates this value empirically through two case studies. In
summary, the research contributions of this paper are:}

\begin{itemize}
\item
  {Proposing a framework of }{five}{~foundational challenges in the
  disinformation research}
\item
  {Developing a methodology to continuously measure and detect
  disinformation relative to a known disinformation narrative based on
  semantic similarity}
\item
  {Evaluating the semantic similarity model on a benchmark to compare
  its similarity scores with}{~humans'}{~similarity scores on the same
  sentence pairs}
\item
  {Interpreting the weaknesses and strengths of these similarity scores
  compared to humans'}
\item
  {Developing a ``tracing tool'' to visualize a timeline of measured
  disinformation}
\item
  {Developing a ``narrative synthesis tool'' to characterize the content
  of detected disinformation}
\item
  {Combining these tools into an accessible and user friendly Tweet
  Narrative Analysis Dashboard}
\item
  {Empirically testing the tracing tool on the target narrative ``The
  2020 election was stolen'' using Donald Trump's tweets, and
  rediscovering studied disinformation patterns in this case study}
\item
  {Empirically testing the tracing tool and the narrative synthesis tool
  on multiple media outlets with the target narrative ``Transgender
  people are harmful to society''}
\end{itemize}

{}

{The research finds that employing a continuous scale metric of
disinformation can be useful to detect, trace, and characterize subtle
patterns of disinformation spread in large amounts of data without
requiring time intensive qualitative analysis, and that this form of
detection is generally aligned with human similarity scores between
text, although some misclassifications occur.}

{}

\section{Background}
{In this section, I establish a conceptual framework useful for
identifying gaps in disinformation research, by categorizing five
fundamental questions in disinformation research. This framework allows
us to narrow the focus and contributions of this paper to disinformation
detection, tracing, and characterization while also recognizing the
potential uses in other fundamental questions. Then, the overview of
LLMs provides the background necessary to understand the NLP methods
that may address gaps in disinformation research by examining LLM
capabilities and limitations. Note that those familiar with the basics of LLMs may want to skip section 2.2. Next, the overview of social media
analysis and quantitative disinformation studies examines the tradeoffs
in current methodologies for detecting and tracing disinformation.
Finally, this allows us to connect the capabilities of LLMs with the
gaps in disinformation detection, tracing, and characterization to
highlight the methodological contributions of the paper.}
\subsection{An Information Disorder Framework}
{This paper revises a framework for disinformation research from Wardle
\& Derakhshan (2017) in order to examine the research contributions made
by the methodologies developed.}{~This paper uses terminology from
Wardle \& Derakhshan's (2017) widely cited overview of the field but
adopts a modified version of their overall framework. The terminology
adopted recognizes the field of ``information disorder'' to encompass
misinformation, disinformation, and malinformation. The framework that
is revised originally consists of three lifecycle phases (creation,
production, and distribution) and three core elements (agent, message,
interpreter) (Wardle \& Derakhshan 2017). }

{This paper will adopt the usage of Wardle \& Derakhshan's (2017) three
types of information disorder, with the following }{definitions: }

\begin{itemize}
\item
  {``Disinformation}{: Information that is false and deliberately
  created to harm a person, social group, organization or }{country}{.}
\item
  {Misinformation}{: Information that is false, but not created with the
  intention of causing harm.}
\item
  {Malinformation}{: Information that is based on reality, used to
  inflict harm on a person, organization or country.''}
\end{itemize}

{}

{However, this paper does not fully adopt their three elements and three
phases of disinformation, but rather expands and refines }{it}{. Below I
develop a revised version of their framework based on }{the ``Five Ws
and How'' approach:}

\begin{itemize}
\item
  {What}{~disinformation is being spread (Detection \&
  Characterization)}
\item
  {Who}{~is spreading it (Attribution)}
\item
  {When}{~and }{where}{~is it being spread (Tracing)}
\item
  {Why}{~is it being spread (Causation)}
\item
  {How}{~is the spread being received and (re)propagated (Impact
  Assessment)}
\item
  {(Bonus) How}{~can it be stopped? (Policy }{Recommendations}{)}
\end{itemize}

{These questions can also be asked }{iteratively over time}{, allowing
for a temporal analysis of disinformation dynamics. This schema adds
explanatory and temporal depth to Wardle \& }{Derakhshan's
original}{~framework. It complements their three elements---Agent (Who),
Message (What), and Interpreter (How)---by incorporating ``When,''
``Where,'' and ``Why,'' dimensions that are critical for a holistic
understanding of disinformation flows. Notably, these central questions
map onto the main challenges of disinformation research: Detection \&
Characterization, Attribution, Tracing, Causation, and Impact
Assessment, as well as the additional layer of Policy
R}{ecommendations}{. }

{Furthermore, }{this approach }{reframes Wardle \& Derakhshan's original
lifecycle model (creation, production, distribution) by asking the
central questions iteratively, thereby offering a more flexible and
analytically rich temporal lens. One reason the original lifecycle model
has not gained widespread adoption may be that disinformation campaigns
frequently operate across all lifecycle phases simultaneously, making
strict phase distinctions less useful in practice.}

{This }{revised framework offered by this paper enables
clearer}{~classification of methodological contributions of this paper.
This framework ties the tracing and narrative synthesis tools to the
central questions of disinformation research concerning
}{what}{~disinformation is being spread and }{when}{~it is spread.
Organizing the field's motivations and methodologies in this way
contributes to a clearer framework for advancing disinformation
research.}
\subsection{Large Language Models' Methodological Contributions}
{Understanding the disinformation detection, tracing, and
characterization contributions developed in this paper requires a
foundational knowledge of large language models, Natural Language
Processing (NLP), and Artificial Intelligence (AI) -- and that is what
this section of the Background aims to provide. The NLP}{~methods in
this paper use open source large language models particularly designed
to compare similarity between texts. This background section will focus
on the basics of large language models, as well as their capabilities
and limitations relevant to this }{paper.  Those familiar with foundational AI and LLM concepts may want to skip to section 2.3. }
\subsubsection{What are Large Language Models}
{LLMs are a specific kind of neural network that leverage a transformer
architecture (Vaswani et al. 2017) to turn natural language into high
dimensional numerical embedding vectors, and often then back into
natural language. In other words, they turn words into big lists of
meaningful numbers and vice versa. These processes are called encoding
and decoding }{respectively}{~(Cho et al., 2014). To encode natural
language, it first is tokenized, which means assigning numbers to
smaller portions of text. Tokenization allows natural language to be
input to the first layer of a large language model (Schmidt et al.
2024). Models are made up of parameters, or weights, which each
represent a multiplication operation that will be applied to their
inputs. Large matrix multiplications using the weights transform the
inputs into outputs the models are trained to generate text, giving
meaning to the numerical representations of the inputs as they pass
through the model (Vaswani et al. 2017).}

{LLMs are often trained by what amounts to filling in the blanks, with
words being ``masked'' and models guessing what word should be in that
place (Yang et al., 2023). Models get progressively better as they are
trained with an algorithm called back propagation (Rumelhart et al.,
1986), which updates the weights. They are trained on massive amounts of
data scraped from the internet. The final output of models can be made
more or less random by changing a hyperparameter (or variable, if you
like) called temperature. Higher temperatures in generative models
increase deviation from the most likely responses (Berger et al., 1996).
A default temperature of 1.0 does not alter the sampling of output
}{tokens}{. }

\subsubsection{LLM Capabilities}
{Though LLMs have applications beyond text---such as code generation and
multimodal AI---in this paper I focus on two core capabilities: encoding
and generating natural language. I focus on these two capabilities
because the tracing tool is founded on encoding two texts to find a
similarity score between them, and the narrative synthesis tool is based
on generating natural language from input text.}

{First, encoding natural language allows us to take a chunk of text and
turn it into a dense vector representation that is imbued with meaning
and some of the context which surrounds it. This is extremely important.
While embeddings for words were available long before LLMs were
developed, LLMs made it possible to represent more surrounding context
in embeddings as well as to embed much longer texts (Vaswani et al.
2017). It is this capability that has spawned the subfield of dense
information retrieval as a branch of information retrieval technology
like search engines (Li et al., 2023). Using LLMs for search can yield
more accurate results that correspond to specific emotions and
subtleties of language not captured by keyword matching and word
frequencies. The utility of LLM-powered search is evidenced by LLMs
dominating leaderboards for reasoning tasks and other benchmarks gauging
mastery of natural language (GLUE Benchmark, n.d.), as well as the
popularity of tools like ChatGPT (Hu, 2023). While some refer to this as
dense information retrieval, in this paper I will use the terminology
``semantic similarity'' to describe the process of comparing cosine
similarity between embedded texts. Thus, for this paper, using the
encoding process and semantic similarity to trace disinformation is
particularly effective because it can more accurately pick up subtle
narratives in various contexts.}

{Second, generating natural language can be used to create new text or
code, answer and follow instructions, summarize inputs and more. In this
paper, LLMs are used to summarize existing text and find the dominant
narratives within it. Summarization and generation is useful in the
context of disinformation to generate new emerging narratives to trace,
to analyze large amounts of detected disinformation, as well as to help
validate the accuracy of tracing methods.}
\subsubsection{LLM Limitations}
{Large language models have many limitations. In particular,
hallucinations and context window size need to be understood to
ethically utilize the tools developed in this paper, as well as a
general recognition of possible other harms and limitations. }

{First, we will examine model hallucinations. Hallucinations -- ~or
responses which are convincing but factually incorrect (Alkaissi \&
McFarlane, 2023) -- ~are inherent to LLMs (Banerjee et al., 2024). This
means that perhaps their greatest flaw is also what makes them useful:
given an input text, they more or less output the average of their
training data's (the Internet's) responses (Heersmink et al., 2024). The
description of this flaw is an oversimplification in many ways, as
different training objectives and sampling schemes greatly affect the
output, but it still demonstrates a broad class of issues facing LLMs.
Hallucinations concern this paper because I generate summaries of
disinformation narratives. Hallucination concerns are discussed further
in the Discussion section.}

{The second important limitation for this paper is context window size.
Although the nature of the matrix multiplication of weights in a
transformer architecture allows for a theoretically limitless input
size, in practice, computers run out of memory to perform such large
operations (Cao et al., 2025). Therefore, models have what is called a
context window, and the size of this determines the largest size of the
inputs. In this case, we encounter this limitation when I generate
summaries of disinformation narratives, and this is again discussed in
further detail later on.}

{LLMs contain many other limitations less directly pertinent to this
paper, and only some will be briefly discussed. As mentioned above,
models are greatly constrained by the quality, quantity, and content of
their training data. The result can be }{racism}{~and unconscious
}{biases}{~in LLMs (Schwartz, 2019; Kotek et al., 2023). These
tendencies and limitations should be kept in mind when using generative
models }{and }{the distilled model used for similarity scores in this
paper -- particularly biases about particular groups potentially imbued
in training data of the models. Possible biases in the narrative
synthesis tool are discussed further in the Discussion. Next, LLMs are
extremely computationally expensive to run. }{This}{~is a growing
environmental concern (Bossert \& Loh, 2025), and as such the uses of
these tools should be scaled within the limits of responsible water and
energy use, and ideally for uses where less computationally expensive
methods are not available or adequate}{.}{~}
\subsection{Social Media Analysis and Quantitative Disinformation Literature Review}

{With an understanding of the technology of LLMs, we now review
disinformation literature and find several methodological challenges
that can be addressed by applying their capabilities. This paper
develops methods that detect, trace, and characterize disinformation on
social media using LLMs to retain nuance and depth of disinformation
narratives detected. Later, the Tweet Narrative Analysis Dashboard
}{applies}{~the tools for temporal frequency analysis, a technique from
social media analysis and disinformation literature. Thus, this review
of}{~how social media can be analyzed, and how it has been in
disinformation literature reveals a lack of depth in characterization
and understanding of nuanced narratives, which the tools developed by
this paper help address.}

\subsubsection{Techniques of Social Media-Based Public Opinion Analysis}
{The subfield of SMPO highlights the various techniques that have been
used in analyzing social media to discern its impact on the world. In a
2021 literature review, }{Dong}{~and Lian outlined five general classes
of analysis, which can be further distilled into four (by combining
spatial and temporal): sentiment analysis, viewpoint analysis, network
user analysis, and spatio-temporal frequency. Sentiment analysis
involves coding the sentiment of words used in social media data,
generally as positive, negative, or neutral. Viewpoint analysis,
frequently interchangeable with ``topic modeling,'' analyzes the most
recurrent themes of messages. Network analysis looks at the links
between information sources on social media. Finally, spatio-temporal
frequency analysis looks at the number of total posts over a given
location and/or timespan. Dong and Lian (2021) show that among these,
temporal, sentiment, and viewpoint analyses are used in a large majority
of SMPO analyses studied. I will describe canonical, large-scale
implementations of the four methods in this review, as well as the
challenges they face in several other studies.}

{El }{Barachi}{~et al. ~(2021) illustrated a common technique for social
media analysis by analyzing over 200,000 tweets related to climate
change posted by Greta Thunberg. This study used frequency analysis
combined with sentiment analysis and location data as a proxy for public
opinion. The study illustrates the usefulness of sentiment analysis by
estimating the public's (strong) support or (strong) opposition to
climate change topics using a type of model called a Bi-directional
LSTM. These models can be used to input text and output classifications
such as support or opposition. The researchers concluded that the
strongest opposition was located in the USA while the strongest support
was from Sweden and the UK. Secondarily, they analyzed the emotions
within Greta Thunberg's tweets over time and correlated this to the
number of retweets and likes. They conclude that ``joy'' and
``discrimination'' emotions yielded the highest engagement while
``anger'' and ``inspiration'' yielded less. Thus, the study demonstrates
how sentiment and frequency analysis can be combined to understand the
reactions and actions of social media users. }

{Large-scale access to digital public forums has also enabled
experimental approaches to measuring changes in public opinion through
social media. Bond et al. (2012) tested the hypothesis that social
influence could drive political mobilization by creating a
61-million-person experiment on the Facebook platform and correlating
their treatments to changes in real-world voting behavior. They did so
by developing features such as buttons suggesting users register to
vote, as well as visuals of Facebook friends who had clicked to register
to vote. Then, they analyzed voter data to validate which users actually
voted. This study shows a type of analysis through experimentation open
to developers at large social media companies. Furthermore, massive
databases collecting and synthesizing data from social and digital media
have enabled new techniques. This includes the study by
Alcántara-Lizárraga and Jima-González (2024), which uses mass-collected
data from the open-source site V-Dem to estimate the effectiveness of
false government information on mass mobilization in Latin American
countries. They employed a variety of statistical methods to estimate
the correlation. They then used Granger causality analysis to test the
direction of causality between these variables. These studies are
representative of the analyses possible on extremely large datasets
using social media to understand the behavior of users or how complex
real world phenomena are influenced by social media.}

{Smaller}{-scale experimentation of social media's effect on public
opinion has also been conducted, specifically in the area of
misinformation. Tokita et al. (2024) studied this by conducting an
experiment with 90 respondents for each of 139 headlines and measuring
the impact on their receptivity to misinformation based on exposure.
They stratified by ideological bias to investigate particular messaging
effects on particular belief structures. Experimentation with survey
responses to measure the effects of social media is one of the closest
to traditional survey techniques used to measure public opinion.}

{Freelon}{, McIlwain, \& Clark (2016) blend spatio-temporal analysis and
viewpoint analysis }{with }{network analysis to quantify the effect of
social media on the Black Lives Matter movement. They focused on three
particular groups involved in social media posts about Black Lives
Matter and used topic modeling (Latent Dirichlet Allocation
specifically) to create these three groups which shared particular
framings. Then, they used spatio-temporal frequency analysis of popular
hashtags and post frequency for all three groups. They used this
analysis to assess the impact on ``elite responses'' which can be viewed
as a proxy for quantifying the power of social media based protest
movements. They estimated the protest power with Granger causality
testing. The unique blending of computational techniques in this study
showed how data and analysis about hashtags and keywords can uncover
influences on powerful people.}

{These}{~studies have shown that a variety of computational techniques
have been employed to analyze public opinion on social media. These
methods are frequently observational, gathering massive amounts of past
data and finding trends. They can be correlated to real-world outcomes
and polls to strengthen the relationship to public opinion. Some
experimental approaches exist, but these are either enacted by platforms
themselves or by small-scale research. Overall, social media's
relationship to public opinion is still being examined by nascent
computational methods combined with traditional measurement techniques.
These techniques are being used in disinformation studies to attribute,
trace, detect, characterize, and measure impacts of disinformation on
social media.}

\subsubsection{Quantitative Disinformation Literature Review}
{Numerous disinformation studies have utilized all four of the
techniques categorized above: sentiment analysis (Osmundsen et al.,
2021; Arcos et al., 2025), spatio-temporal frequency analysis (Field et
al., 2018; Park et al., 2022; Muñoz et al., 2024), viewpoint analysis
(Field et al., 2018; Tuparova et al., 2022), and network analysis
(Starbird et al., 2019; Muñoz et al., 2024; Kennedy et al. 2022). These
techniques allow disinformation researchers to analyze all aspects of
the five foundational challenges of disinformation research but to
varying degrees of specificity and usefulness. This section examines
more closely how these techniques have been used.}

{In their work ``Challenges and Opportunities in Information
Manipulation Detection: An Examination of Wartime Russian }{Media}{'',
Park et al. ~(2022) use keywords to approximate topics and then analyze
the temporal frequency of these topics in various outlets' tweets. They
then employ regression to analyze the framing of the Ukraine war by
state run outlets vs independent tweets. A similar work by }{Field}{~et
al. (2018) uses spatio-temporal analysis of keywords in a Russian
economic newspaper }{Izvestia}{. The researchers uncover a trend of
framing that favors distraction from economic downturn in Russia. Both
of these examples highlight how viewpoint analysis and spatio-temporal
analysis can be used in disinformation research to reveal framings of
contentious world events. }

{The paper ``Divergent Emotional Patterns in Disinformation on Social
Media? An Analysis of Tweets and TikToks about the DANA in Valencia'',
Arcos, Rosso, \& Salaverria (2025) use sentiment analysis to examine the
sentiment of disinformation across TikTok and X, comparing the emotions
on each }{platform}{. Sentiment analysis determined that the nature of
disinformation about flooding in Valencia had higher fear and sadness on
X while TikTok had higher sadness, anger, and disgust levels. They used
a LLM variant of RoBERTa (Liu et al., 2019) for this analysis. Then,
expanding significantly on traditional sentiment analysis techniques,
they helped characterize disinformation by analyzing linguistic features
using an audio based AI model. This paper is a great example of how the
four broad categories of social media analysis techniques are not
all-inclusive and new techniques do not fit cleanly into each category.
It also demonstrates, however, a clear example of how sentiment analysis
as well as other computational characterization techniques have aided
disinformation research.}

{Work by Starbird, Arif, \& Wilson (2019) deepened thinking on
disinformation research frameworks by introducing the notion of
participatory disinformation. Participatory disinformation places
emphasis on network analysis: reactions and collaboration between online
users to spread disinformation, as opposed to specific media outlets and
figureheads as sources of disinformation. To demonstrate a participatory
disinformation analysis, the researchers created network analysis
showing interactions between known Russian ``trolls'' posting about
Black Lives Matter and authentic users retweeting these posts. Another
technique visualizes collaborative work by creating a graph of news
outlets involved in disinformation spread. To do so they find domains
linked by users tweeting specific keywords related to disinformation
narratives, size nodes in a graph based on the number of tweets
containing that domain, and connect nodes based on how many users
tweeted linked to both domains. Then they color each node based on if
the news outlet supported or challenged the disinformation narrative. }

{Beyond individual papers' analyses and methods, some papers have built
open source tools for researchers to use for studying disinformation.
Table 9 in the Appendix provides an overview of relevant tools to
disinformation research and their various uses. What the table reveals
is that although many researchers use similar methods, no standardized
platform or tool has emerged for spatio-temporal, network, or frequency
analysis, as none of the papers reviewed above utilized these }{tools}{.
That does not mean the tools were not used -- or extremely useful to
others -- in any respect, but it does highlight the fragmented and
customized nature of the tools used by disinformation researchers}{.}
\subsubsection{Gaps in Disinformation Literature}
{This}{~paper uses large language models to quantify disinformation
through semantic similarity. The advantage of this approach is that most
of the techniques used to analyze disinformation outlined above can
maintain greater nuance by utilizing a continuous scale rather than a
binary classification of disinformation. For example, the
spatio-temporal method to characterize the framing of Russian economic
disinformation could have used a continuous measure of the distracting
nature of the information being spread, rather than classifying each
data point as about Russia or about the USA. Or, researchers Arif,
Starbird, and Wilson could use a continuous measure of the relatedness
of disinformation being spread about the Black Lives Matter movement to
ensure that accounts were truly spreading this disinformation and root
out false positives and noise in their data. Furthermore, a continuous
metric of similarity might allow the researchers to have a continuous
representation of the political spectrum of their network as opposed to
a binary classification of ``Left-Leaning'' and ``Right-Leaning''. Using
LLMs to measure disinformation in the example of disinformation about
flooding in Valencia could have detected disinformation that their
stringent keyword search might have missed, such as any X user tweeting
about the floods without using the phrase ``DANA''. And, a continuous
scale would offer a degree of how strongly each tweet agrees with a
known disinformation narrative. Possibilities abound. It is this kind of
nuance that a semantic similarity metric using LLMs may offer to
disinformation research.}
\section{Methods}

{Two primary tools were developed for this paper. One tool, the
``narrative synthesis tool'' is for synthesizing emerging narratives on
social media using an LLM for synthesis. The other tool, the ``tracing
tool'' is used to trace specific kinds of information spread from many
posts using semantic similarity scores from a distilled similarity
model.}{~Finally, these two tools are then combined in an interactive
Tweet Narrative Analysis Dashboard for tracing and characterizing
disinformation. The Dashboard is written with the Javascript React
framework, discussed further in the Results section, and available upon
request to the author.}
\subsection{Tool Pipelines}

{The process of the tracing tool follows. Tweets are transformed into
384 dimensional embeddings using a sentence transformer similarity
model. The target narrative to trace is defined and then embedded in the
same space. Then, to produce a similarity score, cosine similarity is
calculated between each tweet and the target narrative embedding.
Finally, tweets are sorted by their similarities for further processing.
}

{}

{The process of the narrative synthesis tool follows. }{To}{~generate
dominant narratives within a body of tweets, the tweets are again
transformed into embeddings using the similarity model. Then, these
embeddings are clustered using K-means clustering based on the
user-selected number of narratives to generate. The number of clusters
significantly varies the results, as different groupings force the model
to focus on different similar aspects of the language. Next, each
cluster of tweets is input to the narrative generation model, with
prompted instructions to return the top two dominant narratives in each
cluster in JSON format. Prompt engineering is performed with a
combination of LangChain's PromptTemplate class and a custom system
prompt. The formatted prompt is injected with the context of the
retrieved tweets (those that are above a set similarity threshold) to
summarize in a manner similar to Retrieval-Augmented Generation
techniques for increasing the accuracy of LLM generation (Lewis et al.,
2021). The formatted prompt is then input to an MLXPipeline using the
narrative generation model, whose output is finally piped to a custom
JSON parser, tailored to the output style of the narrative generation
model. This output is then processed to a Python list, with two dominant
narratives generated per cluster of tweets. Although it would be
intuitive to simply generate one dominant narrative per number of
narratives to generate, I}{~found}{~that allowing the model to be
slightly more verbose by generating multiple produced more accurate
representations of the text. Further experimentation with prompt
engineering could be performed to streamline this process into a more
intuitive one while maintaining the narratives' accuracy. Similarly,
initial experimentation led to setting the MLXPipeline's temperature to
0.9. I chose t}{he temperature}{~based on initial experimentation and it
could be explored more rigorously in future work.}

{}
\subsection{Validation}

{The similarity model is validated for its performance on similarity
tasks using the General Language Understanding Evaluation (GLUE) (Wang
et al., 2019) Semantic Textual Similarity Benchmark (STS-B) (Cer et al.
2017). GLUE is a popular benchmark suite for LLMs, but in this case we
are only interested in evaluating particularly on the STS-B; other
benchmarks within GLUE have already been evaluated and are less relevant
to this particular use case. However, no one has evaluated the
similarity model on STS-B. }

{}

{The STS-B benchmark pairs sentences of varying similarities together,
and asks humans to rank the similarity from 0-5. The rubric of this
ranking system (Cer et al. 2017) is included in the Appendix in }{Table
}{12. There are 1,500 sentence pairs that were ranked by humans. The
similarity model is validated by comparing its cosine similarity scores
(as used in the tracing tool) with a normalized (0-1) STS-B score.}

{}
\subsection{Case Studies}

{In this paper, I will examine two case studies to demonstrate the
capabilities of the methodologies developed. The first case study uses
tweets from Donald Trump during the timeframe 01/01/2020 to 01/01/2021.
This amounts to a total of 12,236 tweets. These tweets will then be used
to trace the similarity to the known disinformation narrative ``The 2020
election was }{stolen}{''. A similarity threshold of 0.45 is applied to
construct a timeline. The election hoax narrative is a well established
falsehood spread by then President Trump to discredit the outcome of the
2020 election (Bowler et al. 2022), which he lost. Then, I generated
three narratives from the tweets in the timeline, with each narrative
generated containing two dominant themes.}

{}

{The second case study examines the emergence of a global narrative that
``Transgender people are harmful to society''. I examine the anti-trans
narrative across four different actors' tweets across the timeframe of
01/01/2024 to 01/01/2025. The similarity threshold is set to 0.38. I
selected this threshold because it reduced the unrelated tweets while
retaining the greatest amount of related tweets during testing. I chose
four actors that include two established left-leaning media outlets and
two right-leaning media outlets. The actors examined are the following:
The New York Times, The Guardian, Fox News, and The Gateway Pundit. In
this timespan, these outlets posted 18,868, 12,513, 31,716, and 19,458
tweets respectively. The Gateway Pundit is categorized as
``hyperpartisan'' and frequently participates in disseminating
disinformation as studied by Starbird, DiResta \& DeButts (2023). It is
included as a useful reference because there are prior studies, such as
that by Starbird, DiResta, and DeButts, on its posting behavior. The
actors chosen here represent a small subset of the possible analyses
that this tool offers, and can be expanded in the future. First, I
perform the narrative tracing ~on all four actors together, then I
compare each with Fox News. This is because Fox News contained by far
the most total tweets, and so comparison served as a sort of relative
maximum frequency.}

{}

{While the first case study provides an opportunity to analyze the
impact of a singular prominent figure's role in spreading disinformation
harmful to democracy, the second case study provides an opportunity to
demonstrate a broader analysis on cultural influencers like established
news outlets, and how these can shape hateful and misinformed
narratives}{.}{~}

{}

{These case studies have a relatively narrow scope and setting compared
to the possible uses of the Tweet Narrative Analysis Dashboard. As
discussed further in the Discussion section 5.4, the case studies in
this paper are focused on social media news outlets and prominent
}{figures}{. The primary reason is the sheer volume of their tweets
available makes it possible to validate the methodologies developed, as
well as the presence of existing literature studying President Trump and
disinformation. The focus on these accounts does not mean, however, that
the same tools do not apply to tracing narratives across multiple
smaller accounts for example. }

{}

{The platform Twitter (X) has been selected as the primary data source
for analysis but it should be emphasized that the methods of analysis
and tools developed in this paper can be used for any textual data. It
could also be applied to video transcripts. For the first case study,
however, it was important to select a well documented disinformation
narrative and platform, so that the focus of the case study was
validating the new methodology for tracing and generating narratives
rather than application of the methods to unknown contexts. I selected
Twitter because of the existing disinformation literature about the 2020
election hoax on this platform as well as the availability of archived
tweets by Donald Trump during this time. Data used for this case study
are from Brown (2016). Data for Case Study 2 are from Junkipedia
(2024).}

{}

\section{Results}
\subsection{Similarity Score Validation on the GLUE STS-B}
{In the following, the similarity model is evaluated on the GLUE
Semantic Textual Similarity Benchmark and the results are evaluated
using the Pearson correlation coefficient. The GLUE STS-B benchmark was
developed by Cer et }{al}{. (2017) and can be used to evaluate semantic
search systems. }
\subsubsection{GLUE STS-B Results}
{The similarity model scores a 0.8696 Pearson correlation, which shows
significant correlation between human similarity scores and model
similarity scores. The major differences are represented by Figure 1
which shows that the model's average error is 0.1383, and these errors
primarily concentrate around similarity scores of 0.2 - 0.4. }

{}
Model interpretation is performed, comparing examples of most similar
model predictions and most dissimilar predictions, as well as examining
examples from each quartile of the model's error range. Additional
examples of model similarity scores for each quartile can be found in
the Appendix and were used to help interpret the causes of the model's
error.

\begin{figure}[H]
  \centering
  \title{\textbf{Figure 1: Correlation between Human and Model Similarity Scores}}
  \includegraphics[width=\textwidth]{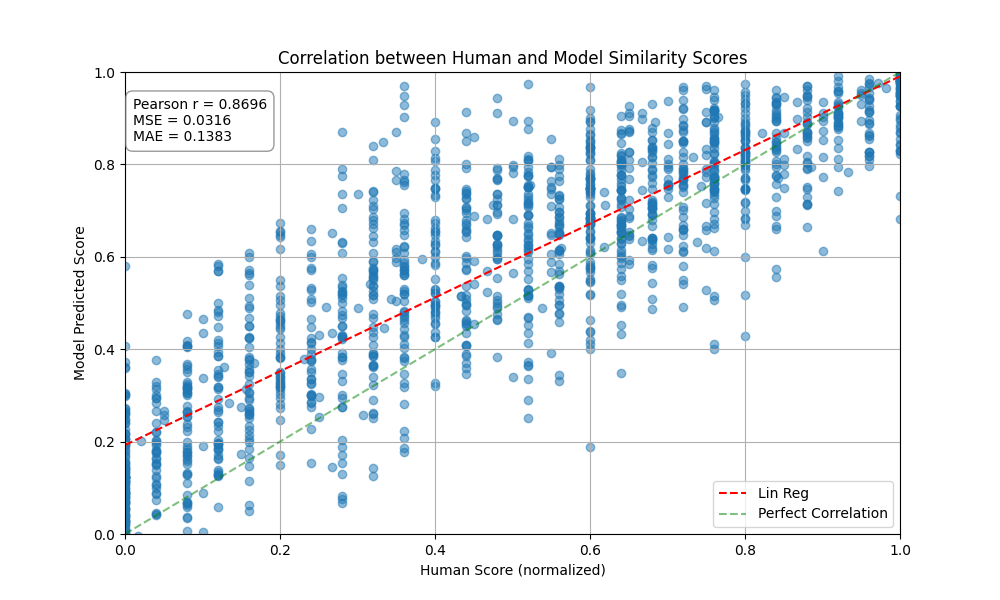}
  \caption{We plot the similarity model's predicted similarity between 1,500 pairs of sentences in the GLUE STS-B against humans' scores of the similarity between those sentence pairs. The plot reveals that the model's similarity scores and human's similarity scores are highly correlated, with 0.8696 Pearson's correlation and a mean average error of 0.1383.}
  \label{fig:similarity-scatter}
\end{figure}

\begin{figure}[H]
  \centering
  \title{\textbf{Figure 2: Plotting Model Residuals Based on Human Similarity Score}}
  \includegraphics[width=\textwidth]{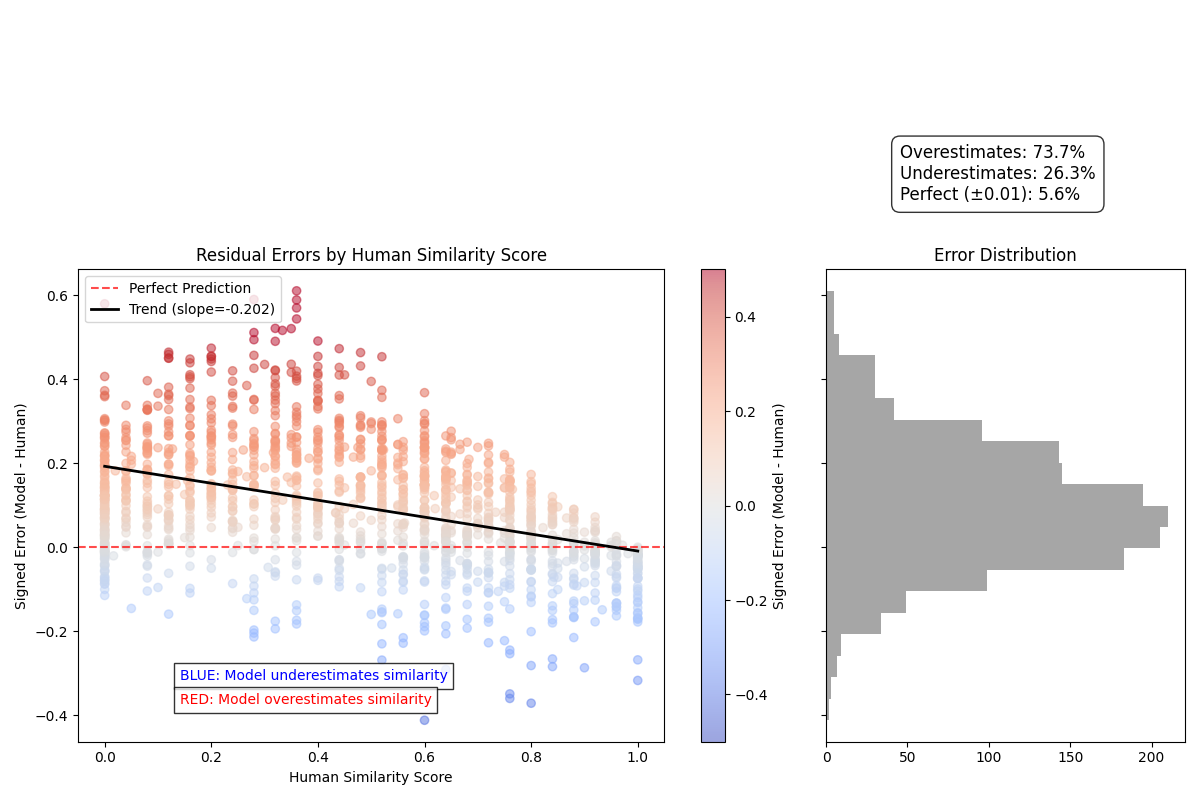}
  \caption{The figure examines the signed error of the model's
similarity scores compared to humans' scores, revealing that the model
tends to overestimate the similarity of sentences compared to humans,
and particularly so on sentences that humans rank as more dissimilar. It
also shows that the distribution of the model's error is centered around
its mean signed error in a bell shaped curve, so that large errors for
either overestimated or underestimated similarity are rarer than small
deviations.}
  \label{fig:residuals}
\end{figure}

\subsubsection{GLUE STS-B Examples by Human Score Range}\label{h.rmcczs8sr8oo}

\textbf{Table 1: Summary of Model Classifications and Errors on GLUE STS-B}
\begin{longtable}{p{2.2cm}p{2cm}p{2cm}p{2cm}p{2cm}p{2cm}}
\toprule
Approximate Human Similarity Score & Average Absolute Error & Average Signed Error & Number of Examples & \% Over-estimated & \% Under-estimated \\
\midrule
\endfirsthead

\multicolumn{6}{c}{{\textbf {Table \thetable\ continued from previous page}}} \\
\toprule
Approximate Human Similarity Score & Average Absolute Error & Average Signed Error & Number of Examples & \% Over-estimated & \% Under-estimated \\
\midrule
\endhead

\midrule
\multicolumn{6}{r}{{Continued on next page}} \\
\endfoot

\bottomrule
\endlastfoot

0.0 & 0.1424 & 0.1276 & 291 & 84.5 & 15.5 \\
0.25 & 0.2043 & 0.1834 & 306 & 88.9 & 11.1 \\
0.5 & 0.1585 & 0.1277 & 363 & 82.1 & 17.9 \\
0.75 & 0.0942 & 0.0371 & 379 & 67.0 & 33.0 \\
1.0 & 0.0635 & -0.0497 & 161 & 21.7 & 78.3 \\

\end{longtable}

\vspace{-0.5cm}
\noindent\textbf{Table 1:} The table quantifies the model's tendency to overestimate the similarities based on bracketed human similarity scores. It also shows that the model has the highest absolute and signed error in the lower half of human similarity scores, from around 0.5 to 0.0 on the human scale.
\\
{}

{We can now interpret several examples of the model's lowest error
similarity scores and highest error similarity scores for each bracket
of human similarity scores. The interpretation of model scores provides
an intuition for the reasoning behind the model's differences in
similarity scores, and the areas where it agrees most with human
similarity scores. Tables 2 - 6 show examples of model scores and
errors, on sentence pairs grouped by human scores ranging around 0.0,
0.25, 0.5, 0.75, and 1.0.}
\\
\\
\textbf{\noindent{Table 2: Human Score Examples $\approx 0.0$}}

\small
\begin{longtable}{p{0.8cm}p{2.2cm}p{2.2cm}p{1.33cm}p{1.33cm}p{0.8cm}p{3.0cm}}
\toprule
Type &  Sent. 1 & Sent. 2 & Human Score & Model Score & Error & Analysis \\
\midrule
\normalfont
\endfirsthead

\multicolumn{7}{c}{{\textbf Table continued from previous page}} \\
\toprule 
Type &  Sent. 1 & Sent. 2 & Human Score & Model Score & Error & Analysis \\
\midrule
\endhead

\midrule
\multicolumn{7}{r}{{Continued on next page}} \\
\endfoot

\bottomrule
\endlastfoot

Best & The fruits should be eaten with lemon juice in order to prevent oxidation in your stomach. & Three dogs growling on one another & 0.0000 & 0.0007 & 0.0007 & The model almost perfectly matches the human score. \\
\midrule
Worst & 3 killed, 4 injured in Los Angeles shootings & Five killed in Saudi Arabia shooting & 0.0000 & 0.5794 & 0.5794 & The model overestimates similarity because although they share similar topics, according to STS-B guidelines, this should not score more than 0.4. \\

\end{longtable}
\normalsize

\vspace{0.5cm}

\textbf{\noindent{Table 3: Human Score Examples $\approx 0.25$}}

\small
\begin{longtable}{p{0.8cm}p{2.2cm}p{2.2cm}p{1.33cm}p{1.33cm}p{0.8cm}p{3.0cm}}
\toprule
Type &  Sent. 1 & Sent. 2 & Human Score & Model Score & Error & Analysis \\
\midrule
\endfirsthead

\multicolumn{7}{c}{{\textbf Table continued from previous page}} \\
\toprule 
Type &  Sent. 1 & Sent. 2 & Human Score & Model Score & Error & Analysis \\
\midrule
\endhead

\midrule
\multicolumn{7}{r}{{Continued on next page}} \\
\endfoot

\bottomrule
\endlastfoot

Best & A man is riding a bicycle on a dirt path. & Two dogs running along dirt path. & 0.3200 & 0.3232 & 0.0032 & The model is extremely concurrent with human scores. The sentence swaps the subject and slightly alters the verb but has a very similar setting. \\
\midrule
Worst & The Note's Must-Reads for Friday, December 6, 2013 & The Note's Must-Reads for Friday, July 12, 2013 & 0.3600 & 0.9703 & 0.6103 & The model identifies shared topics and sentence structure despite temporal differences. The STS-B rubric favors maximum similarity of 0.4 since the sentences are technically ``not equivalent''. \\

\end{longtable}
\normalsize

\vspace{0.5cm}

\textbf{\noindent{Table 4: Human Score Examples $\approx 0.5$}}

\small
\begin{longtable}{p{0.8cm}p{2.2cm}p{2.2cm}p{1.33cm}p{1.33cm}p{0.8cm}p{3.0cm}}
\toprule
Type &  Sent. 1 & Sent. 2 & Human Score & Model Score & Error & Analysis \\
\midrule
\endfirsthead

\multicolumn{7}{c}{{\textbf {Table continued from previous page}}} \\
\toprule 
Type &  Sent. 1 & Sent. 2 & Human Score & Model Score & Error & Analysis \\
\midrule
\endhead

\midrule
\multicolumn{7}{r}{{Continued on next page}} \\
\endfoot

\bottomrule
\endlastfoot

Best & A group of people eat at a table outside. & A group of elderly people pose around a dining table. & 0.5200 & 0.5172 & 0.0028 & The model recognizes the similar subject and setting but also the difference of details. \\
\midrule
Worst & 10 Things to Know for Wednesday & 10 Things to Know for Thursday & 0.4000 & 0.8908 & 0.4908 & Again, the model identifies shared sentence structure despite temporal differences. The STS-B rubric favors maximum similarity of 0.4 since the sentences are technically ``not equivalent''. \\

\end{longtable}
\normalsize

\vspace{0.5cm}

\textbf{\noindent{Table 5: Human Score Examples $\approx 0.75$}}
\small
\begin{longtable}{p{0.8cm}p{2.2cm}p{2.2cm}p{1.33cm}p{1.33cm}p{0.8cm}p{3.0cm}}
\toprule
Type &  Sent. 1 & Sent. 2 & Human Score & Model Score & Error & Analysis \\
\midrule
\endfirsthead

\multicolumn{7}{c}{{\textbf Table continued from previous page}} \\
\toprule 
Type &  Sent. 1 & Sent. 2 & Human Score & Model Score & Error & Analysis \\
\midrule
\endhead

\midrule
\multicolumn{7}{r}{{Continued on next page}} \\
\endfoot

\bottomrule
\endlastfoot

Best & A young blonde girl wearing a smile and a bicycle helmet. & A young girl wearing a bike helmet with a bicycle in the background. & 0.7600 & 0.7599 & 0.0001 & The model recognizes the similar subject and actions but also slight differences. \\
\midrule
Worst & Higher courts have ruled that the tablets broke the constitutional separation of church and state. & The federal courts have ruled that the monument violates the constitutional ban against state-established religion. & 0.8000 & 0.4285 & 0.3715 & The model fails to recognize the similarity between the objects in these sentences, perhaps due to unusual and formal prose. \\

\end{longtable}
\normalsize

\vspace{0.5cm}

\textbf{\noindent{Table 6: Human Score Examples $\approx 1.0$}}

\small
\begin{longtable}{p{0.8cm}p{2.2cm}p{2.2cm}p{1.33cm}p{1.33cm}p{0.8cm}p{3.0cm}}
\toprule
Type &  Sent. 1 & Sent. 2 & Human Score & Model Score & Error & Analysis \\
\midrule
\endfirsthead

\multicolumn{7}{c}{{\textbf Table continued from previous page}} \\
\toprule 
Type &  Sent. 1 & Sent. 2 & Human Score & Model Score & Error & Analysis \\
\midrule
\endhead

\midrule
\multicolumn{7}{r}{{Continued on next page}} \\
\endfoot

\bottomrule
\endlastfoot

Best & Colorado Governor Visits School Shooting Victim & Colorado governor visits school shooting victim & 1.0000 & 1.0000 & 0.0000 & The model accurately sees these as perfectly semantically similar despite syntactic differences. \\
\midrule
Worst & A dog jogs through the grass. & A dog trots through the grass. & 1.0000 & 0.6827 & 0.3173 & The model inaccurately focuses on subtle linguistic differences that reduce the similarity score compared to human judgement. \\

\end{longtable}
\normalsize
\subsubsection{GLUE STS-B Model Interpretation: Residual Error Analysis}

{The similarity model from sentence transformers tends to overestimate
the similarity of sentences that humans find more dissimilar, while
performing very well on extremely similar and middle similarity
sentences. The average error of the model on human similarity scores of
0 is 0.1424, while at 0.5 is 0.1585 and at 1 is 0.0635. The model
generally overestimates low similarity scores (signed error at 0:
0.1276) and slightly underestimates very high similarity scores (signed
error at 1: -0.0497). Therefore, the model will likely contain more
false positives than false negatives, and in the context of tracing
disinformation, this means the model might flag legitimate differences
as similar content. This overestimation bias should be considered when
setting similarity thresholds for applications like detecting
disinformation variants. We can also see through examining examples of
the worst predictions that the model struggles with nonsensical
sentences, and can significantly underestimate similarity due to
grammatical errors, as seen in the 1.0 human score worst prediction. The
exact prevalence of underestimation due to grammatical errors is not
clear, but the average error of the model relative to the human score is
shown in Table 1. We can also see that the model tends to overestimate
the similarity of sentences with differing dates but similar structure.
Although this behavior could be desirable in certain contexts, the
discrepancy could be due to the STS-B's description of 0.2 similarity
ratings as "The two sentences are not equivalent, but are on the same
topic." (Appendix, Table 12).}
\subsubsection{Similarity Score Interpretation}

{Given that the model scores a high Pearson coefficient to human
similarity scores and we have examined the areas where scores differ the
most, we can now provide a table to summarize an intuitive
interpretation of the similarity scores that the model provides. The
lower middle to middle ranges had the highest average signed error and
are the most difficult to interpret.}
\textbf{Table 7: Interpretations of Model Score Ranges}

\small
\begin{longtable}{p{2.5cm}p{5cm}p{5.5cm}}
\toprule
Model Score Range & Example (model score) & Interpretation \\
\midrule
\endfirsthead

\multicolumn{3}{c}{\textit{Table continued from previous page}} \\
\toprule 
Model Score Range & Example (model score) & Interpretation \\
\midrule
\endhead

\midrule
\multicolumn{3}{r}{{Continued on next page}} \\
\endfoot

\bottomrule
\endlastfoot

0.75 - 1.0 & (0.8896) The man without a shirt is jumping. 

The man jumping is not wearing a shirt. & The same sentence, or extremely close. A sentence with a typo generally achieves a score of 1.0 paired with the same sentence without a typo. The same sentence differing only by dates mentioned will generally be in this bracket. \\
\midrule

0.5 - 0.75 & (0.5794) 3 killed, 4 injured in Los Angeles shootings

Five killed in Saudi Arabia shooting & May share sentence structure, subject(s), object(s), events, and details, but with slightly differing phrasing and/or meaning. Importantly, texts sharing subjects and objects but with opposite sentiment are often grouped in this bracket. \\
\midrule

0.25 - 0.5 & (0.3232) A man is riding a bicycle on a dirt path.

Two dogs running along dirt path. & Lower range is mostly dissimilar in meaning though may share syntax. Upper range indicates some shared meaning. \\
\midrule

0.0 - 0.25 & (0.1528) The academic year does start around September in the USA and I think most European countries.

I would not accelerate things, to avoid getting worse grades that you want. & Highly dissimilar, possibly completely unrelated. Upper range may share similar details at most. \\

\end{longtable}
\normalsize
\subsection{Features of the Tweet Narrative Analysis Dashboard}
{I developed the Tweet Narrative Analysis Dashboard to combine the
functionalities of the tracing tool and the narrative synthesis tool
into one simple dashboard for tracing and characterizing disinformation
in a continuously quantitative manner. The following walks through the
features developed.}

{}

{In Figure 3, the user}{~}{can see the first feature allows us to add
multiple datasets to analyze and graph simultaneously using the Add
Dataset button. To reset the number of datasets analyzed, refresh the
page.}

{}

{Second, the user can select which dataset to analyze from the datasets
found in the tweets folder of the code repository (available upon
request of the author).}

{}

{This dataset has several formatting requirements: it must be a csv
file, it must contain Twitter post body content in column
``post\_body\_text'', contain a timestamp in column ``published\_at'',
and, if embedded content of tweets is to be analyzed, contain embedded
content in column ``EmbeddedContentText''. }

{}

{Next, the user can input the target narrative to trace across their
dataset. The target narrative should be natural language and is most
effective when formatted as a complete sentence rather than keywords or
phrases.}

{}

{Several features control the parameters of analysis: we can select a
timeframe and the minimum similarity threshold }{to graph}{. The
similarity timeline graphs only tweets that are equal or greater than
the threshold in cosine similarity with the target narrative. The
similarity threshold can be set to 0 to consider all tweets in the
dataset. }

{}

{Finally, after tracing a narrative from the dataset, narratives can be
generated from the subset of the dataset graphed. For each bulleted
narrative that is generated, two dominant themes of each narrative will
be presented.}

{}

{Both the tracing tool and the narrative synthesis tool can take time. A
dataset of 50,000 tweets can take over an hour to trace on a standard
laptop with 16GB of RAM using Apple's M1 processor with an MLX model.}

{}

{The Tweet Narrative Analysis Dashboard demo is available upon request
to the author.}
\subsection{Case Study 1: Tracing Election Hoax Disinformation in Trump's Tweets}
\begin{figure}[H]
  \centering
  \title{\textbf{Figure 3: Analysis of the Election Hoax Narrative}}
  \includegraphics[width=0.8\textwidth]{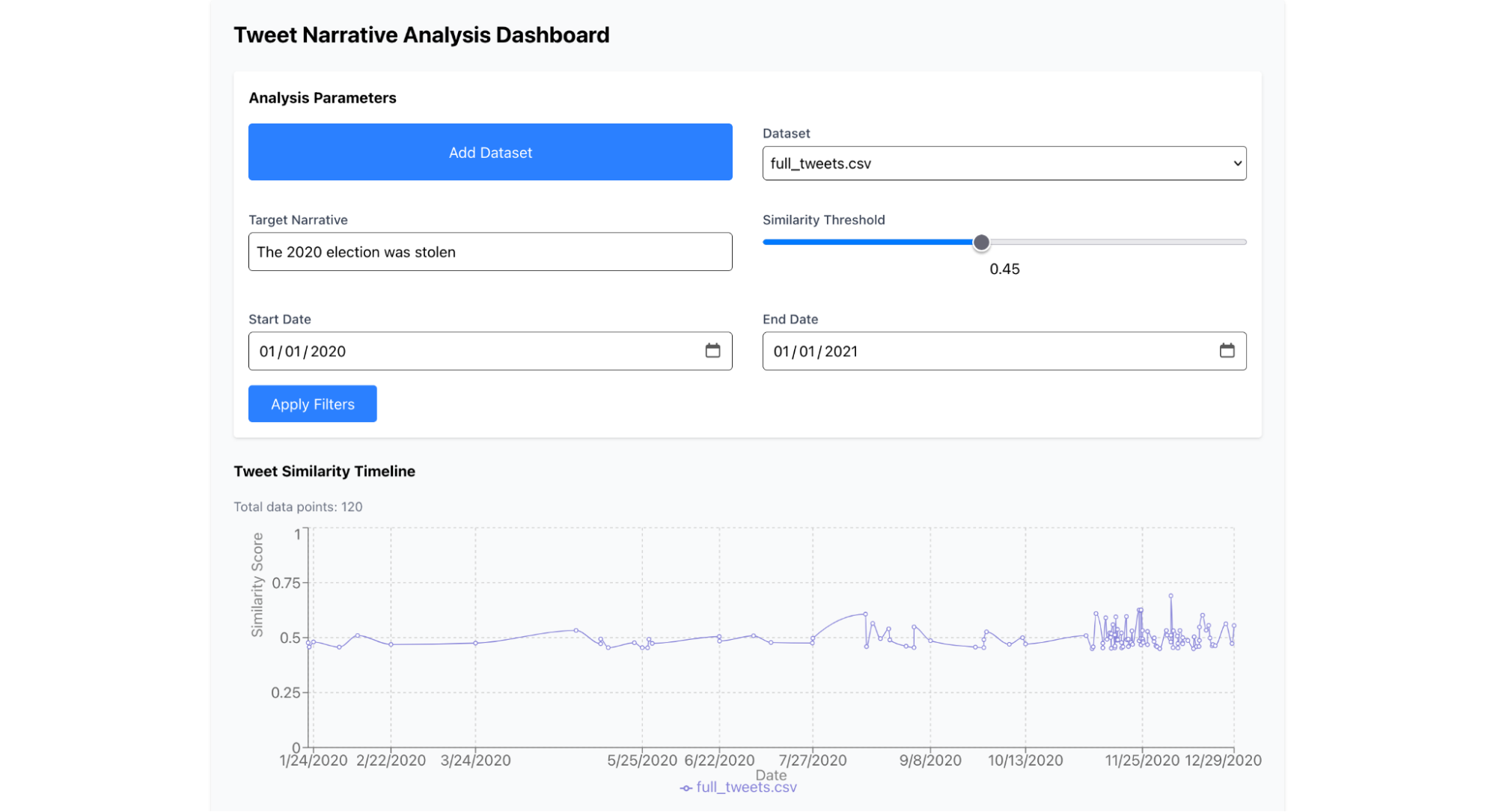}
  
  \vspace{0.3cm}
  
  \includegraphics[width=0.8\textwidth]{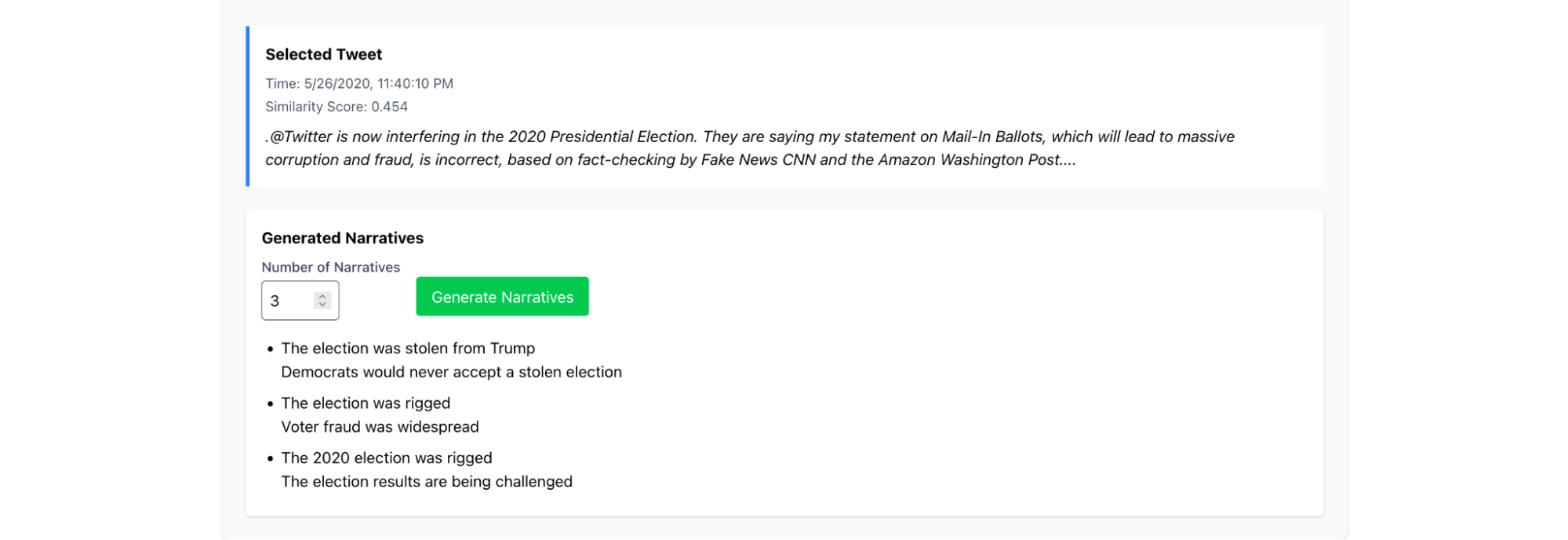}
  \\
  \caption{Analysis of the Election Hoax Narrative. Results from the Tweet Narrative Analysis Dashboard which show the frequency of tweets similar to the target narrative ``The 2020 election was stolen'' above a threshold of 0.45 similarity, and the generated narratives among those tweets.}
  \label{fig:election-analysis}
\end{figure}

{}

{The tracing tool shows a pattern of higher activity of disinformation
during November 2020, after Donald Trump lost the election. }{However,
analyzing a longer timeframe, }{we can also see the seeds of doubt sowed
prior to the election}{, helping to make the claims of voter fraud
ambiguous to his millions of followers}{. These seeds of doubt can be
seen clearly in the tweet selected which references ``Mail-in Ballots''
leading to ``massive corruption and fraud''. }

{}

{Generating three narratives in this case confirms the accuracy of the
tracing tool, as the six different themes for the three separate
narratives are different phrasings of the original target narrative. The
similarity of the generated narratives to the target narrative is useful
validation of the accuracy of the trace. In more complex cases, for
example using the target narrative ``Russia is a U.S. ally'', or in
cases where the similarity threshold is lower, the generation feature
can be useful for characterizing the different dimensions of spreading
disinformation. This is examined further in Case Study 2.}

\subsection{Case Study 2: Tracing Anti-Trans Disinformation in Mass Media}
\begin{figure}[H]
  \centering
  \title{\textbf{Figure 4: Comparing Anti-Trans Tweet Similarity Across Four News Outlets}}
  \includegraphics[width=\textwidth]{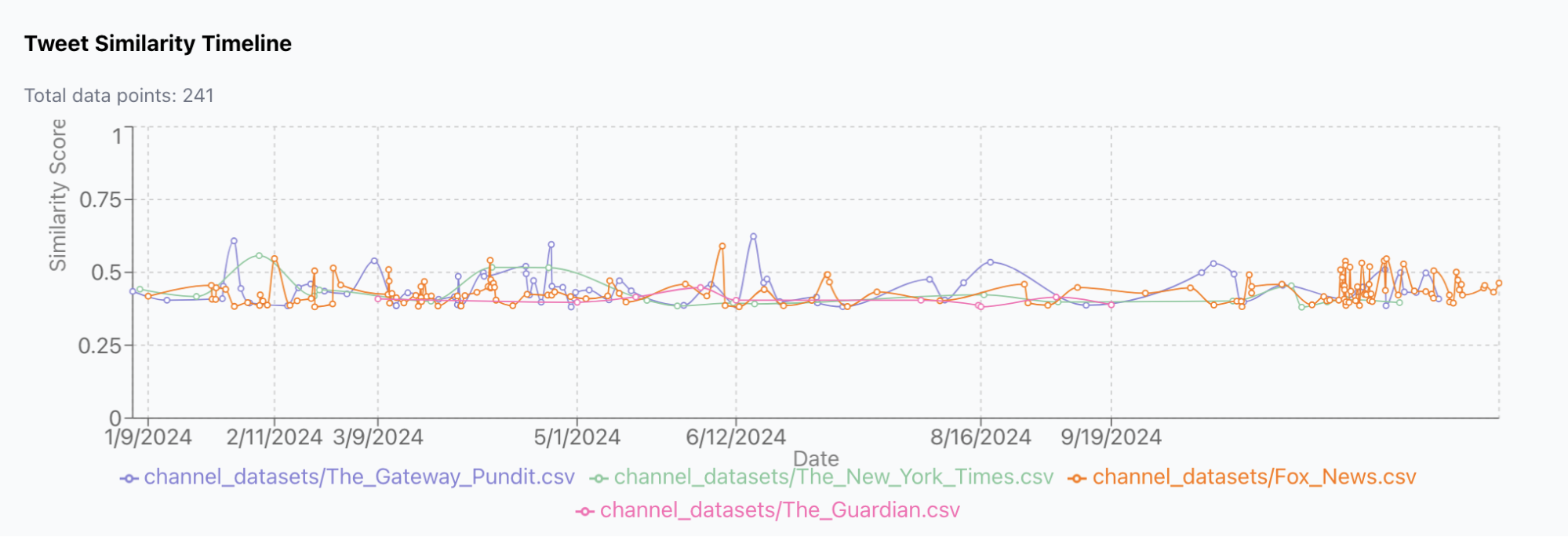}
  \caption{A combined timeline of all four news outlets compared in Case Study 2, graphed for the narrative ``Transgender people are harmful to society'' with the similarity threshold of 0.38.}
  \label{fig:anti-trans-combined}
\end{figure}

\begin{figure}[H]
  \centering
  \title{\textbf{Figure 5: Comparing Anti-Trans Tweet Similarity With Fox News}}
  \includegraphics[width=\textwidth]{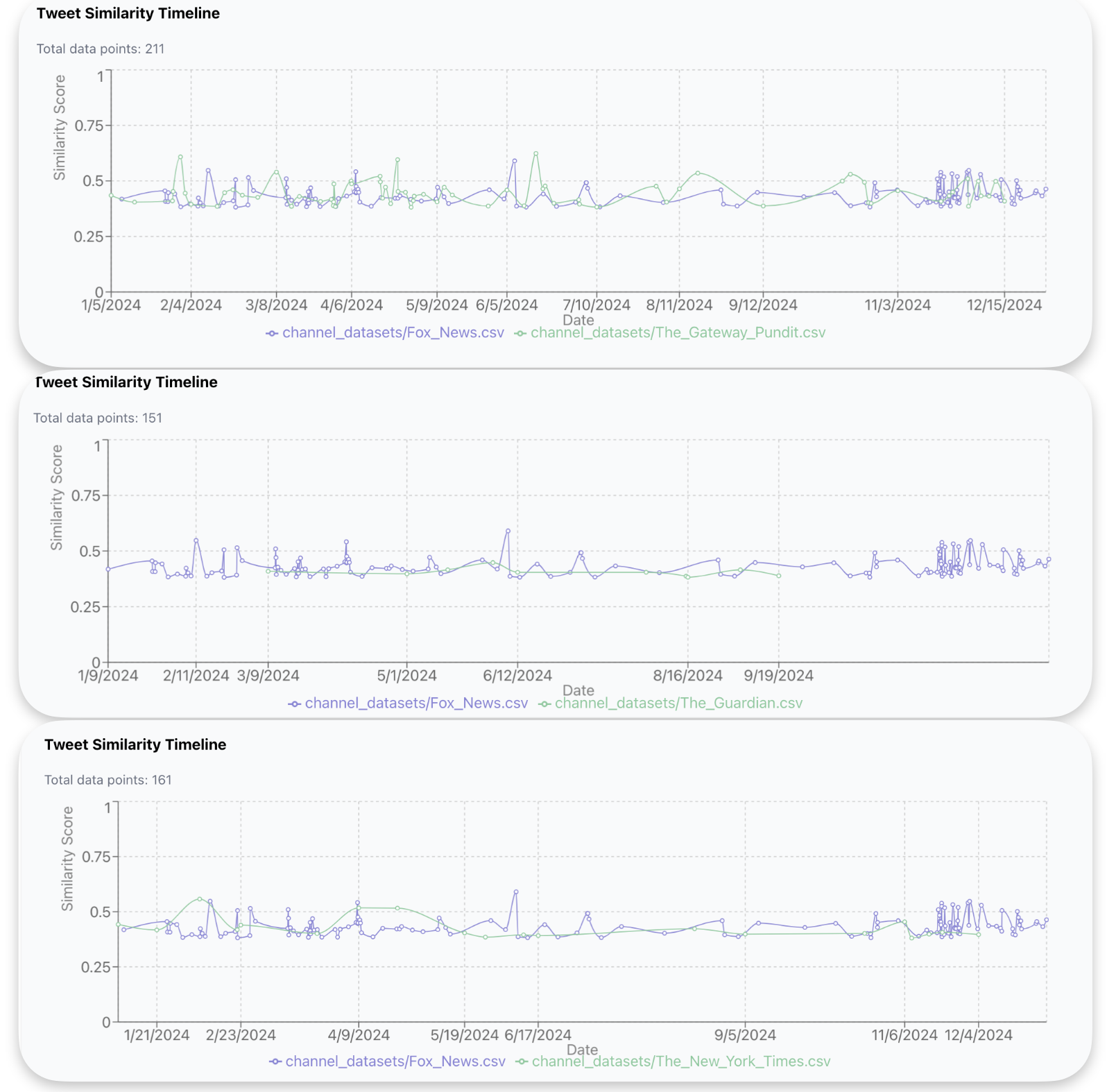}
  \caption{The figure shows the plots of The Gateway Pundit, The Guardian, and The New York Times' similarity timelines from Figure 4, plotted with Fox News' timeline for reference.}
  \label{fig:anti-trans-fox-comparison}
\end{figure}

\begin{figure}[H]
  \centering
  \title{\textbf{Figure 6: Individual News Outlets' Anti-Trans Tweet Similarities}}
  \includegraphics[width=\textwidth]{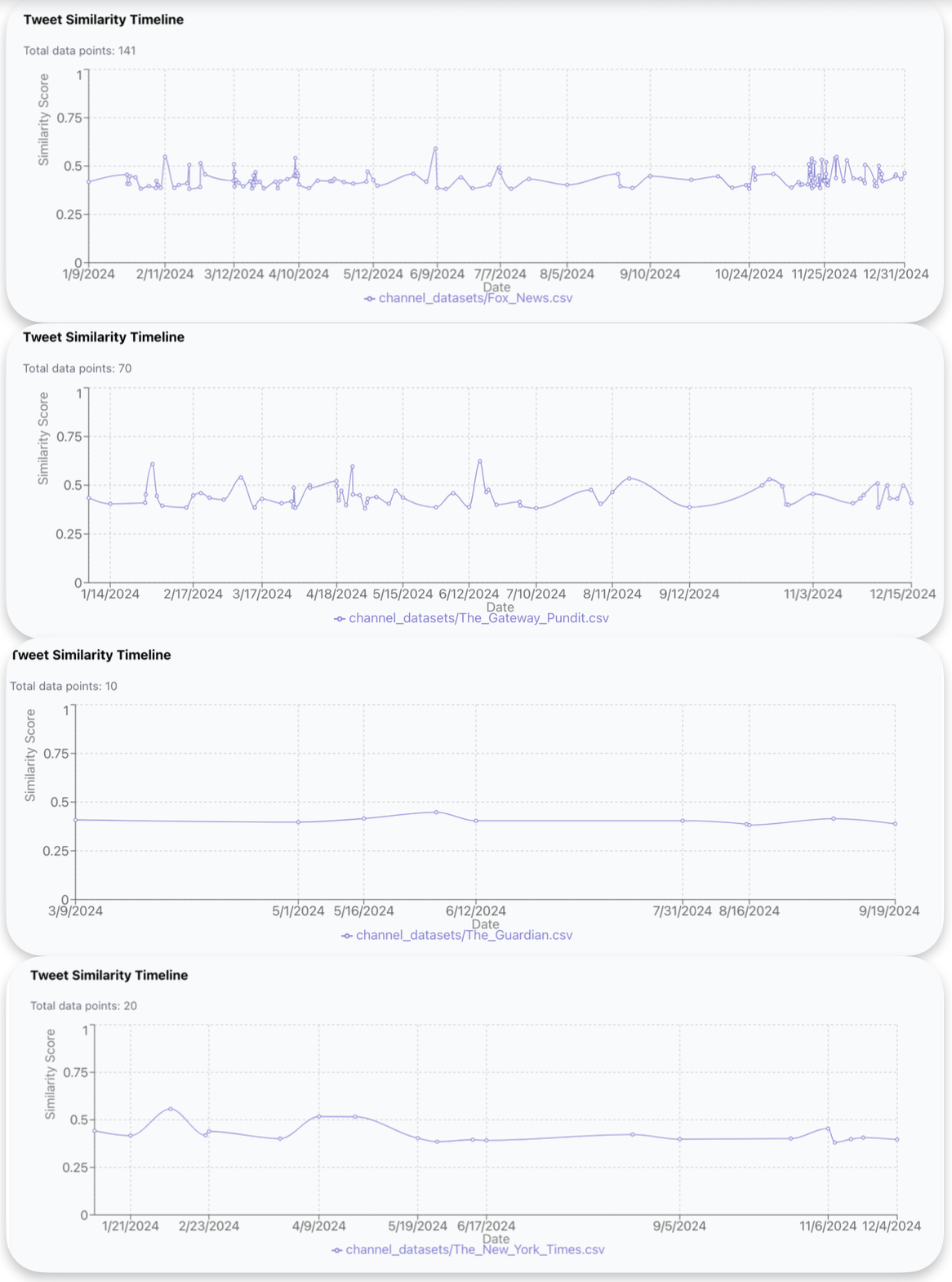}
  \caption{The figure shows the Fox News, The Gateway Pundit, The Guardian, and The New York Times' similarity timelines from Figure 4, separately plotted.}
  \label{fig:anti-trans-individual}
\end{figure}

\begin{figure}[H]
  \centering
  \title{\textbf{Figure 7: Generated Narratives within Anti-Trans Tweets by Outlet}}
  \includegraphics[width=\textwidth]{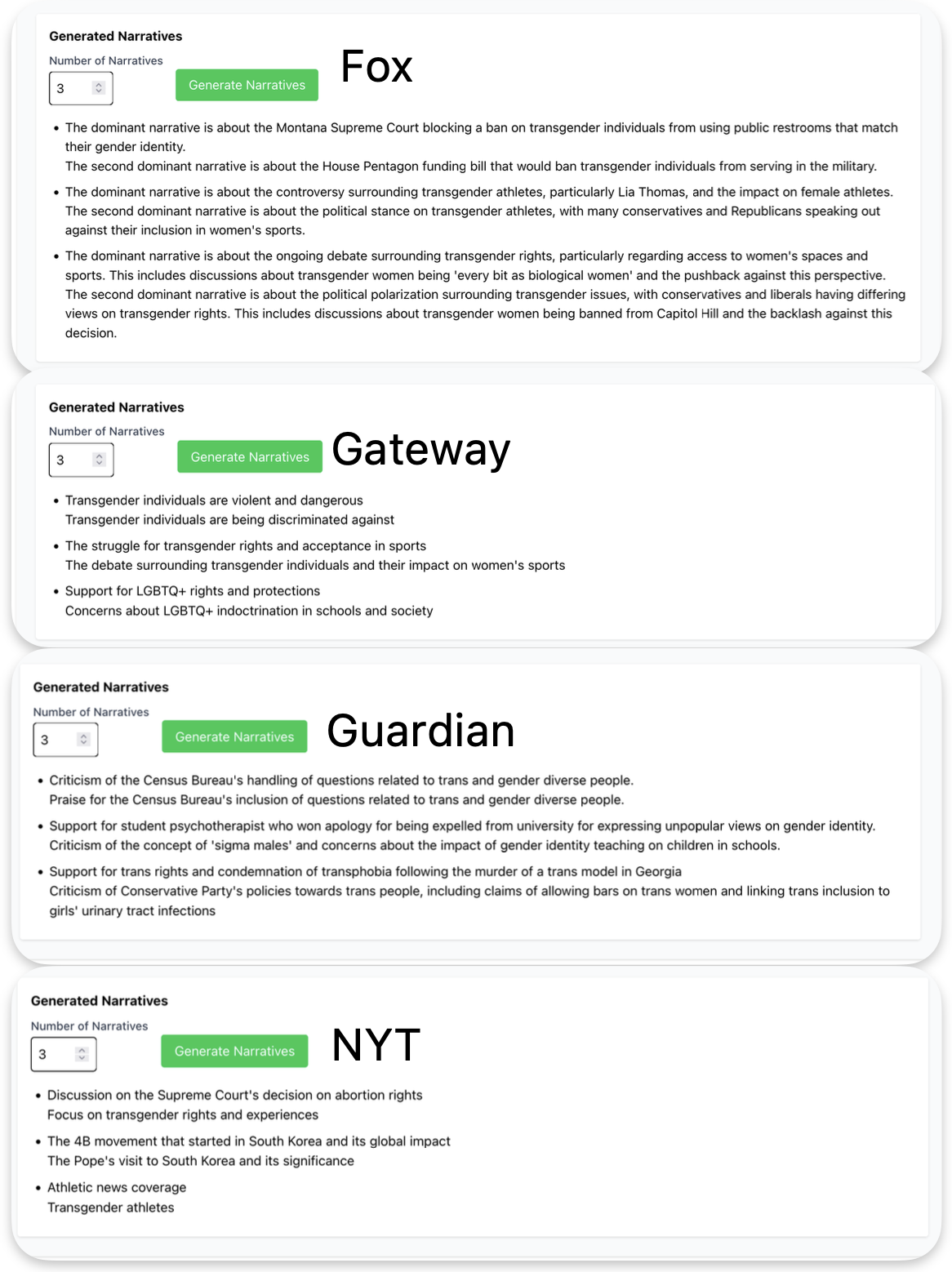}
  \caption{The figure shows the narratives generated from the tweets in the timelines from Figure 4 for Fox News, The Gateway Pundit, The Guardian, and The New York Times. A typed and perhaps more readable version of this figure is available in the Appendix in Table 13.}
  \label{fig:anti-trans-narratives}
\end{figure}
\section{Discussion}
\subsection{Validating the Methodology}

{I validated t}{he similarity scoring mechanism developed for this paper
by careful comparison to the GLUE STS-B benchmark. The model had a
strong correlation with human judgement, with a 0.8696 Pearson
correlation. Further examination of the differences in human and model
scores revealed subjective and occasionally more intuitive scores given
by the model than humans. Examination also revealed that the model
occasionally made important mistakes in similarity scores, but the
strong Pearson coefficient and overall low average error (0.1383) gives
confidence that these mistakes are infrequent. Therefore, the tracing
tool, whose only LLM is the similarity model, can be used to trace
disinformation across social media, though with cautious interpretation
of any singular data point. }

{}

{I did not further validate the LLM used for the narrative synthesis
tool in this paper. Instead, its base model was evaluated on numerous
standard AI commonsense and reasoning benchmarks by its developer
Mistral AI (Mistral AI, 2024). In these benchmarks it outperforms many
other popular mid sized LLM models. One important note is that although
the base model was evaluated on these benchmarks, the
}{instruction-finetuned and }{MLX-converted}{~version of the model
used in this paper has not been. Finetuned models exhibit only
``marginal performance gaps'' on Bioasq and Natural Questions
}{datasets}{~(Barnett et al., 2023). Furthermore, baseline models could
easily be swapped }{in to}{~the narrative synthesis tool if more compute
power was available.}

{}

{Now, having confidence in the validity of the models used in the
methodology of this paper, we can turn to its contributions to
disinformation research.}

\subsection{Case Study 1: Research Contributions}

{One of the important contributions made by this paper is putting
disinformation into a continuous quantitative realm, where its
disinformation can be approximately measured in relation to human
designed and studied narratives, rather than classified as either
disinformation or not. Measuring disinformation in this relative way
allows for automated detection while still capturing the subtle ways
that disinformation can mix with truth. Continuous measurement of
disinformation expands the possibilities for the core challenges of
characterization, attribution, tracing, and assessing impact of
disinformation. For example, in the first case study, we are able to
trace the known disinformation narrative ``The 2020 election was
stolen'' and create a timeline of tweets containing this disinformation,
and then approximate the ``amount'' of disinformation by scoring the
similarity to this narrative. The timeline reveals a pattern of higher
activity of disinformation beginning in November 2020, but also reveals
consistent doubt sewn about election integrity in months leading up to
the election. This analysis is consistent with a European
}{Parliament}{~(2021) finding that, ``Trump framed the debate about the
outcome of the presidential election perhaps anticipating that he
would lose months, even years ago,'' but that after his loss, he
``reheated the claims of voter fraud}

{and appeared to encourage protests.'' The tracing tool developed by
this paper allows for a significantly more nuanced understanding of how
and when these narratives were propagated because rather than a temporal
analysis based on the number of disinformation-classified tweets over
time, we can inspect how similar the narrative is over time. Finally,
this methodology is repeatable, transparent, and inspectable by other
researchers as the code and models are available.}

{~ }

{The tracing tool was also useful for disinformation detection by using
natural language rather than keywords. Detecting disinformation with
keyword search, by definition, does not consider the semantic context
and may lead to misclassifications. The contributions made through
semantic similarity as a metric move towards a semantically holistic
approach, where the intent and context of specific narratives can be
considered during detection. Search terms (target narratives) can be
whole strings of natural language rather than specific keywords,
deepening the capabilities of analysis and disinformation tracing. Case
Study 1 demonstrates this by capturing tweets which contain keywords
that may or may not indicate disinformation, and quantifying their
similarity. For example, on 11/16/2020, Trump tweets, ``I won the
election!'' and this scores 0.489 similarity. This simple tweet would
only contain the possible keyword ``election'', and the presence of this
keyword alone cannot classify something as disinformation (not to
mention the problems already discussed with such binary
classifications). Another example is the more complex tweet from
05/01/2020 which reads ``RT @RealJamesWoods: These were left in the
lobby of an apartment building. They are unsecured ballots ripe for
`harvesting' by crooked Demo\ldots'' (the text was cut off by Twitter's
140 character limit at the time). The similarity score for this example
is 0.455. In this example, the main keyword present would be ``ballots''
and maybe if the keyword search was especially thorough, ``insecure'',
but it is unlikely that a researcher would use the word
``}{unsecure}{'', or its even more uncommon adjective form
``unsecured''. Again, the presence of the word ``ballot'' is not a good
indicator of disinformation. So we can see the limits of keyword
searches in this example, where the natural variety of language can make
it difficult to capture all the possible related tweets containing
disinformation. Furthermore, the process of developing keyword searches
may lend itself to confirmation bias, as researchers iteratively
construct strings to find tweets they already consider disinformation
(Kennedy et al. 2022). The tracing tool could be used to complement
existing processes for detection and compared to check for biases. Using
semantic similarity helps detect disinformation that resists keyword
search and can even act as a tool to test for confirmation bias that may
come from the process of designing keyword searches.}

{}
\subsection{Case Study 2: Research Contributions}

{The tracing tool in Case Study 2 highlights how disinformation can be
compared and tracked across multiple outlets, and how spatio-temporal
and network analyses can benefit from the nuance of a continuous scale
measure of disinformation. We can see that Fox News posts by far the
most similar content to the target narrative ``Transgender people are
harmful to society''. Initially, this might be expected given that their
total number of tweets is at a 1.6/1 ratio to the Gateway Pundit, 2.5/1
compared to the Guardian and 1.7/1 compared to the New York Times.
However, upon closer examination, the number of tweets above the
threshold is at a 2.0/1 ratio compared to The Gateway Pundit, 14.1/1
ratio compared to The Guardian, and a 7.1/1 ratio compared to the New
York Times. }

\textbf{Table 8: Comparison of Fox News Anti$-$Trans Tweet Frequency}

\small
\begin{longtable}{p{2.5cm}p{2cm}p{3cm}p{3cm}p{2.5cm}}
\toprule
Outlet & Total Tweets & Ratio of Total Tweets to Fox News & Tweets $>$ 0.38 Similarity Ratio to Fox News & Overall Anti-trans Tweet Rate \\
\midrule
\endfirsthead

\multicolumn{5}{c}{\textit{Table continued from previous page}} \\
\toprule 
Outlet & Total Tweets & Ratio of Total Tweets to Fox News & Tweets > 0.38 Similarity Ratio to Fox News & Overall Anti-trans Tweet Rate \\
\midrule
\endhead

\midrule
\multicolumn{5}{r}{{Continued on next page}} \\
\endfoot

\bottomrule
\endlastfoot

Fox News & 31,716 & 1:1 & 1:1 & 4.41e-3 \\
\midrule
The Gateway Pundit & 19,458 & 1.6:1 & 2.5:1 & 3.60e-3 \\
\midrule
The Guardian & 12,513 & 2.5:1 & 14.1:1 & 0.799e-3 \\
\midrule
The New York Times & 18,868 & 1.7:1 & 7.1:1 & 1.05e-3 \\

\end{longtable}
\normalsize

\vspace{-0.5cm}
\noindent Table 8: The table shows the total tweets in each outlet analyzed in Case Study 2 for the target narrative ``Transgender people are harmful to society'' as well as their ratios of total tweets compared to Fox News and their total tweets above similarity threshold 0.38 compared to that of Fox News.

{Fox had 141 tweets above the 0.38 similarity threshold out of 31716
total. Graphing each outlet separately in Figure 6 displays the obvious
difference in frequency of anti-trans similar tweets between the right
and left leaning outlets. The ratios with respect to left leaning
outlets are far higher than the ratio of total tweets to Fox News in the
given timespan. The tool also reveals a pattern of heightened focus on
this narrative prior to the 2024 election for Fox News. There was also a
heightened frequency during the period of March 2024 and May 2024 for
both Fox News and The Gateway Pundit. However, neither The New York
Times nor The Guardian had relatively higher similar tweets during this
time. The reasons for this could be further examined by using the
tracing tool to narrow the timeframe and either generating or manually
examining the tweets for specific attributes of the narratives.
Observations from the timeline about disinformation similarity frequency
could be combined with expertise from political science researchers and
international studies to understand the global context driving these
observed trends in narrative shift. Another interesting observation
revealed in Figure 5 shows that there may be some correlation between
Fox News and the Gateway Pundit's frequency of similar anti-trans
tweets, with similar spikes around June 2024 and lack of spikes in the
months thereafter, for example. An interesting direction would be to
perform causality testing on the two }{timeseries}{~to discern the
magnitude and direction of this possible correlation and/or causality.
The ability to quickly graph these two analyses with the tracing tool
enabled this insight. Another observation from Figure 5 is that the
number of datapoints above the 0.5 similarity threshold is much higher
for both Fox and The Gateway Pundit than for either The New York Times
(which has 3) or The Guardian (which has none). In this way, the
continuous scale allows us to visualize the strength of the similarity
to the target narrative and discern potential differences in the
characteristics of the similar tweets. Overall, the tracing tool can
visualize more nuance than traditional temporal analyses through a
continuous metric of similarity to a target narrative.}

{}

{The narrative synthesis tool can characterize the particular attributes
and content when analyzing large amounts of disinformation data.
However, this case study used a significantly downsized model for
computational purposes, and the performance is not as desirable as that
of larger models might be. That being said, we can see in narratives
generated in Figure 7 (or Table 13 in the Appendix) that the tweets from
The Gateway Pundit have an emphasis on violence caused by trans people.
This is certainly not a model hallucination. Here just a few examples:}

{``ABSOLUTELY SICK: Transgender `Vampire' Sexually Assaults Disabled
Minor \textbar{} Beyond the Headlines via @gatewaypundit'' (Similarity
score: 0.521)}

{``Judge Who Put Transgender Child Rapist in Women's Prison Nominated to
U.S. District Court by Joe Biden via @gatewaypundit'' (Similarity score:
0.387)}

{``WATCH: Male Student Who Identifies as Transgender Injures THREE Girls
During Basketball Game -- Causing Opposing Team to Forfeit via
@gatewaypundit'' (Similarity score: 0.460)}

{Interestingly, the paired dominant narrative is ``Transgender
individuals are being discriminated against''. This is also not
necessarily a hallucination, as several of The Gateway Pundit's tweets
would seem neutral to trans rights, when removed from the context of
their other posts. For example: ``British Lawn Tennis Association Bans
Transgender Women From Most Female Tournaments''. The generated
narratives thus show the importance of contextualizing any single
narrative generated among the multiple others generated. It also
demonstrates how mixing partisan headlines with more neutral ones can be
employed by this outlet as a disinformation strategy. Overall, the
narrative synthesis tool provided useful insights into the
characteristics of The Gateway Pundit's similar tweets.}

{}

{The narratives synthesized from The Guardian's tweets are hyperspecific
and demonstrate a weakness of the tool. Because The Guardian only had
ten total tweets above the 0.38 similarity threshold and I clustered
them into three groups in order to generate narratives, the model has
potentially only a single tweet in a given cluster. This causes the
hyperspecificity of the generated narratives, and is less useful as a
tool for detecting broad trends. The 20 tweets from the New York Times
also suffer from this weakness, focusing on a particular event in South
Korea for one narrative generated. However, we can still compare this
with the narratives generated from The Gateway Pundit and see that The
Guardian does not frame transgender people as violent or dangerous.
Thus, although the narrative synthesis tool is most useful on mid-sized
datasets (several hundred datapoints), it can still be useful to compare
broad differences in disinformation characteristics.}

{}

{The narrative synthesis tool is also useful to discern the
characteristics of the anti-trans tweets from Fox News. While the output
is verbose, it is accurate in capturing Fox News' focus on transgender
people in restrooms, serving in the military, participating in sports,
and being considered women. This observation alone is not a particularly
interesting insight, as this mostly summarizes the contentious issues
around transgender people in America during 2024, but it does help to
validate the accuracy of the tool. Furthermore, the narratives generated
also help to capture the stance, such as in the generated clause, ``with
many conservatives and Republicans speaking out against their inclusion
in women's sports'', or the statement, ``This includes discussions about
transgender women being `every bit as biological women', and the
pushback against this perspective''. Thus, the narrative synthesis tool
shows utility in uncovering a nuanced stance in Fox News' tweets similar
to the target narrative that transgender people are harmful to society.
}
\subsection{Limitations and Future Directions}
{The introduction of the methodology in this study is limited in many
respects and has much that can be improved for future use in the field
of disinformation. }

{}

{One of the major limitations is this study's focus on the Twitter / X
platform. Disinformation is often spread across platforms, and, even
more complexly, can originate in fringe forums and platforms. Thus,
expanding the novel methodology developed here to support multiple
platforms is essential for more holistic addressing every foundational
challenge in disinformation research. Second, disinformation spread is
participatory (Starbird et al., 2019). The case studies in this paper do
not focus on participatory disinformation but rather on tracing
disinformation in prominent public figures. This focus was not meant to
demonstrate that that is the most effective way to study disinformation
spread, but rather a way to focus on a high impact area with limited
access to data. Studying participatory disinformation requires large
scraping of comments and interactions. Gathering Twitter/X data is now
financially difficult on Twitter/X due to policy changes enacted in
February }{2023}{, which installed a paywall for previously free Twitter
API access (Gotfredsen, 2023). Future research could expand to study
participatory disinformation on other platforms, such as Reddit which
offers free API access. Thus, researchers with access to more data or
funds would be able to use these tools to analyze disinformation using a
participatory framework.}

{}

{The narrative synthesis tool has several known concerns. One particular
concern is the potential for hallucination. Hallucination did not appear
to occur while testing this model, as the model is primed with
information to summarize directly in its prompts, and is not asked about
information outside of this domain. However, it is possible that if
given sparse data to summarize, or given topics which were heavily
discussed in the model's training data, it may hallucinate narratives
that were not present in the data presented in the prompt. Model
hallucination is an area for further testing and validation.}

{}

{Another known concern of the narrative synthesis tool is trying to
summarize too many tweets and overflowing the model's context window.
The model used in this paper has a context window of 8k
}{tokens}{~(Mistral AI, 2023). A very rough calculation which estimates
four characters per token based on testing with OpenAI's
}{tokenizer}{~(Open AI, n.d.) amounts to a context window of
approximately 128 total 250-character tweets maximum per narrative
generated. Context window size is a consideration for the user to take
into account when uploading datasets and using the narrative synthesis
tool.}

{}

{Also, the narrative synthesis tool could be validated more rigorously
for use in the field of disinformation by comparing generated narratives
to expert coding, as well as comparing the contents of the clustering
mechanism the expert coding. A comparison to human-summarized narratives
would help identify possible biases in the model toward particular
figures or groups. Understanding training data bias in the narrative
synthesis tool ~is particularly important if we consider the case
studies in this paper: because of the frequently partisan nature of
disinformation and possible bias in training data, disinformation could
be embedded in LLM training, and LLMs could be biased against detecting
particular kinds of disinformation. In Case Study 1, this is accounted
for by generating narratives that recreated the target election hoax
narrative, which was known disinformation. The similarity of the
generated narratives and the target narrative confirmed that bias had
not prevented the narrative synthesis tool's output from matching expert
studies on the same topic. However, in the second case study, as its
nature was more exploratory, such confirmation is lacking. Thus, further
comparison with expert coding is a welcome direction for future
research.}

{}

{A limitation of the sentence similarity model is that it occasionally
finds opinions which are on the same topic but with opposite sentiment
to be of a relatively high similarity (usually around 0.5). For example,
tweets by the New York Times in support of transgender people, but
mentioning `harmful' -- a keyword that was present in the target
narrative -- scored particularly high in similarity, presumably because
both ``trans'' and ``harmful'' were present in the tweet. These kinds of
false positives are rare, but can be addressed with qualitative
characterization, or with further processing using LLMs. }
\begin{figure}[htbp]
  \centering
  \title{\textbf{Figure 8: High Similarity of a Tweet with Opposite Sentiment to the Target Narrative}}
  \includegraphics[width=\textwidth]{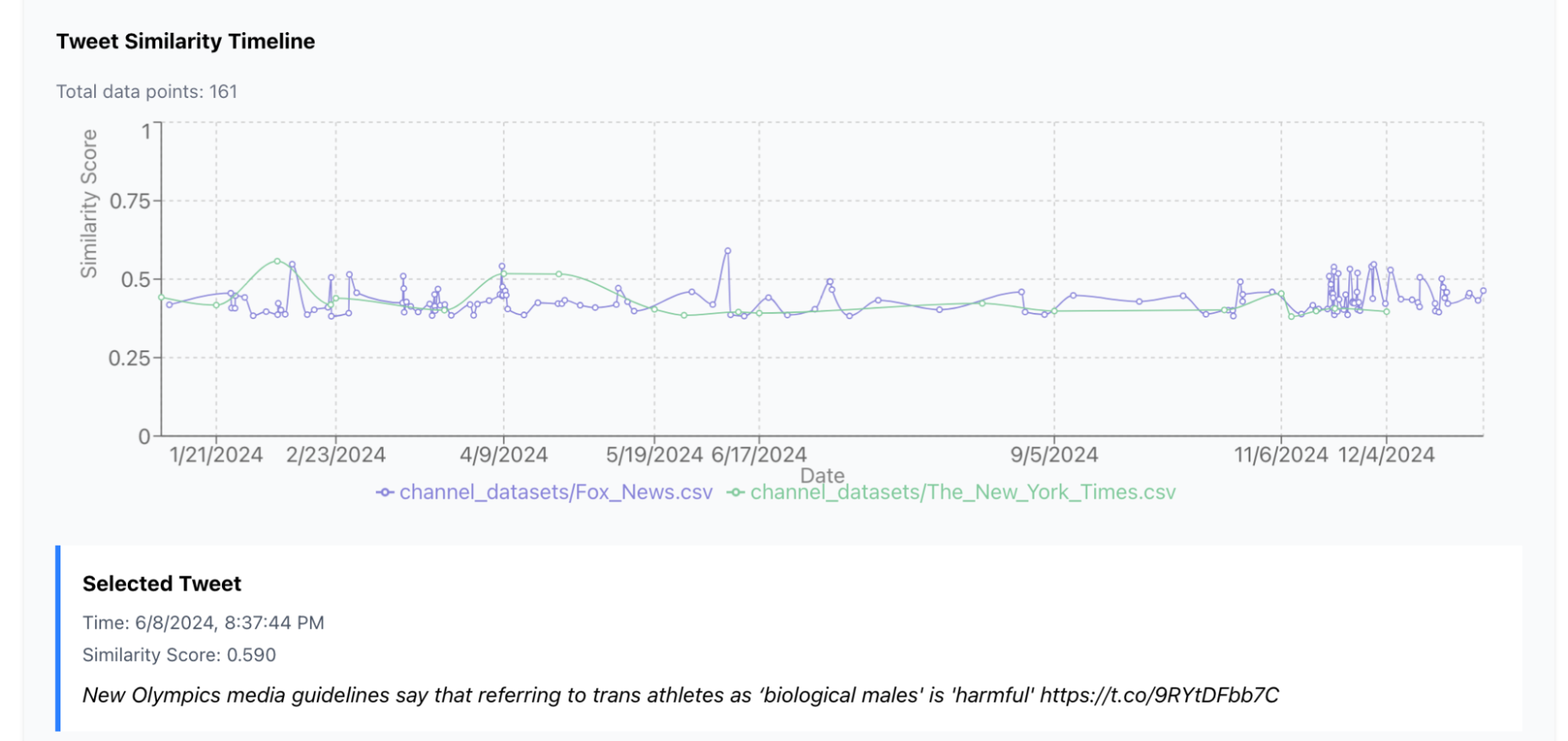}
  \caption{High Similarity of a Tweet with Opposite Sentiment to the Target Narrative. The figure displays a `misclassification', selecting a tweet which scores a high similarity (0.590) with the second case study narrative, but is not relatively more aligned with the meaning behind the target narrative.}
  \label{fig:misclassification}
\end{figure}

{For example, this limitation can be overcome by performing sentiment
analysis on sentences and flipping the polarity of the similarity score
according to the agreement of the sentiment. I experimentally flipped
the polarity of the similarity score using LLM-based stance detection,
but the results are not included in this paper, as I did not validate
the method for accuracy compared to human judgement. For a qualitative
approach, using the tracing tool to narrow the search for disinformation
narratives and then manually coding the results can still cut down on
hundreds of thousands of data points, in a way aligned with human
judgement to a high degree (see section 4.1). Therefore, while the
similarity score does align with human judgement for most cases,
misclassifications occur and can be overcome with careful analysis.}

{}

{One very promising future direction is to expand the analysis to longer
text. I hypothesize that longer text would give the similarity scores
greater accuracy, as the context is more nuanced and detailed than 250
(or 140) character limited tweets. Another important direction would be
to expand this approach to images, given that images can be a
particularly potent medium of disinformation (Wardle \& Derakhshan,
2017). To do so would require the use of visual embeddings, social media
data replete with images, and benchmarking on image-focused datasets.
Transferring the methodology developed in this paper to images and
longer text is achievable with existing technologies. }

{}

{This technology poses several concerning ethical considerations. First,
the uses of the Tweet Narrative Analysis Dashboard are not limited to
the study or prevention of disinformation. The Dashboard can be useful
to analyze social media in many different ways, including commercial
interests that serve more to fuel information disorder than to combat
it. Furthermore, it could potentially be used in an offensive (as
opposed to defensive) manner to find and amplify the most effective
disinformation narratives. Don't do that. That is why access is
available by request. I won't grant you access if you plan to do that.}

{}

{Second, like the process of crafting keyword searches, the tracing
tool's use can be prone to confirmation bias. In Case Study 1, this
paper only looked for disinformation in a dataset where it was already
present, in order to validate the usefulness of the methodology
developed. However, the practice of looking for trends where they are
already suspected could lead to biased analyses. The tracing tool can be
used in a systematic manner to avoid bias. This was demonstrated in Case
Study 2 where multiple data sources were compared from across multiple
ideological viewpoints and media sources. Like any tool, the cognitive
biases of the user must be considered as it is used.}{~}

\section{Conclusion}

{Disinformation research has an extremely wide array of quantitative and
mixed detection methods at its disposal, but these tools often make
simplifications in order to handle more data that detract from the
quality of the characterization of disinformation. This paper offers a
method to apply a continuous scale measurement of disinformation based
on semantic similarity to known disinformation narratives. This
methodology allows researchers to replace keyword based methods of
disinformation detection with a method that can better understand
context and semantics of whole sentences and bodies of text.
Furthermore, this continuity of semantic similarity means that
disinformation can be analyzed in a way that acknowledges the
non-binary, gray area between truth and falsehood that is often a
hallmark of disinformation. Finally, the tools built to achieve this are
accessible with an easy to use user interface, with the hope that these
methodological contributions will be useful for the field.}

\section{Appendix}
\subsection{Disinformation Research Tool Analysis}

{The RAND Corporation assembled a database of disinformation tools that
have been released publicly (RAND Corporation, 2022). A review of these
tools finds that a majority are intended to fight misinformation, and
resemble variations of fact checking tools. However, several of these
tools pertain to the present research and an overview will be provided
here. Several relevant tools from outside of this database are also
compiled here.}
\\
\textbf{Table 9: Overview of Open Disinformation Tools}

\small
\begin{longtable}{p{4cm}p{9cm}}
\toprule
Tool & Function \\
\midrule
\endfirsthead

\multicolumn{2}{c}{\textit{Table continued from previous page}} \\
\toprule 
Tool & Function \\
\midrule
\endhead

\midrule
\multicolumn{2}{r}{{Continued on next page}} \\
\endfoot

\bottomrule
\endlastfoot

Social Data Search (Benzoni, 2024) & Searchable database which automatically labels, monitors, and updates with ``outputs from sources that we can directly attribute to the Russian, Chinese, or Iranian governments or their various news and information channels.'' \\
\midrule

Information Laundromat (Information Laundromat, 2025) & Made by the European Media and Information Fund, this is best used to find a singular common source of one narrative or type of content among known state sponsored media sites and known fake news sources. Checks website meta information like domain certificates, hosting providers, etc and identifies networks of similar sites. Can also rate similarity of contents and titles of search results. \\
\midrule

Ad Observer Chrome Extension (Ad Observer, n.d.) & Browser Extension which tracks data about ads on Facebook that you see while highlighting those it classifies as political. \\
\midrule

Who Targets Me (Who Targets Me, 2018) & Browser extension used by 100,000+ to reveal the political leanings of ads users see. Also visualizes political ad spending on Meta and Google. \\
\midrule

Bot Sentinel (Bot Sentinel, 2024) & Deprecated February 2023: Bot Sentinel used to analyze Twitter accounts and scored their probability of violating Twitter guidelines using a classification model that was trained by, ``searching for accounts that were repeatedly violating Twitter rules and we trained our model to classify accounts similar to the accounts we identified as `problematic.''' It was prone to false positives on non-English accounts. \\
\midrule

Botometer (OSoMe, n.d.-a) & Deprecated: Twitter Bot Detector trained a bot detecting AI model based on account metadata. Contains several datasets of annotated bot accounts through the ``Bot Repository''. \\
\midrule

Hoaxy 2.0 (OSoMe, n.d.-b) & Visualizes the spread of information on Bluesky, or Twitter if you have paid API access (at least \$100/month). More details: ``Hoaxy visualizes temporal trends and diffusion networks. Temporal trends plot the cumulative number of posts over time. The user can zoom in on any time interval. Diffusion networks display how posts spread from person to person. Each node is an account and two nodes are connected if a post is passed between those two accounts. Larger nodes represent more influential accounts. The color of a connection indicates the type of post: reposts, replies, quotes, or mentions.'' \\
\midrule

SMILE (National Center for Supercomputing Applications, 2025) & Automatic scraping of YouTube and Reddit data (Twitter is no longer accessible) for academic researchers. Also provides automated NLP analysis techniques like topic modeling, sentiment analysis, text classification, and more. \\
\midrule

EU Vs. Disinformation (European External Action Service, n.d.) & Journalists trace and compile Russian narratives in the media, namely through RT and Sputnik, and provide alternative narratives and facts to counter them. \\
\midrule

Meta Ad Library API (Meta, 2025) & Searchable database of Facebook ads that are both active and inactive, supplying their target audience countries, estimated audience size, and more information. \\
\midrule

OSoMeNet (OSoMe, 2025) & Network analysis of reposts and interactions on various social media platforms based on keyword search. \\

\end{longtable}
\noindent{
Table 9: Tools for tracking partisan ads, detecting bots, and
gathering state sponsored media abound. Some display masterful data
visualizations to show the spread through social media. However, none of
these tools can trace a disinformation narrative providing the
continuous temporal analysis and characterization contributed by this
paper.}
\normalsize

\subsection{GLUE STS-B Examples by Model Error Quartile}
\textbf{Table 10: Model Examples on the GLUE STS-B Grouped by Error Quartile}

\small
\begin{longtable}{p{1.5cm}p{4cm}p{4cm}p{1.5cm}p{1.5cm}p{1.2cm}}
\toprule
Type & Sentence 1 & Sentence 2 & Human Score & Model Score & Error \\
\midrule
\endfirsthead

\multicolumn{6}{c}{\textit{Table continued from previous page}} \\
\toprule 
Type & Sentence 1 & Sentence 2 & Human Score & Model Score & Error \\
\midrule
\endhead

\midrule
\multicolumn{6}{r}{{Continued on next page}} \\
\endfoot

\bottomrule
\endlastfoot

Excellent & The dogs are chasing a cat. & The dogs are chasing a black cat. & 0.8800 & 0.8696 & 0.0104 \\
\midrule

Excellent & The man without a shirt is jumping. & The man jumping is not wearing a shirt. & 0.9200 & 0.8896 & 0.0304 \\
\midrule

Excellent & But JT was careful to clarify that it was ``not certain about the outcome of the discussion at this moment''. & ``However, we are not certain about the outcome of the discussion at this moment.'' & 0.6400 & 0.6834 & 0.0434 \\
\midrule

Good & A person is peeling shrimp. & A person is preparing shrimp. & 0.7200 & 0.7722 & 0.0522 \\
\midrule

Good & As mentioned in the other comments, ANOVA is problematic when mixing types of predictor variables. & I like to think of multitasking as rapid task switching. & 0.0000 & 0.1081 & 0.1081 \\
\midrule

Good & Sirius carries National Public Radio, although it doesn't include popular shows such as ``All Things Considered'' and ``Morning Edition.'' & Sirius recently began carrying National Public Radio, a deal pooh-poohed by XM because it doesn't include popular shows like All Things Considered and Morning Edition. & 0.6800 & 0.7892 & 0.1092 \\
\midrule

Fair & The academic year does start around September in the USA and I think most European countries. & I would not accelerate things, to avoid getting worse grades that you want. & 0.0400 & 0.1528 & 0.1128 \\
\midrule

Fair & There is a older man near a window. & A boy is near some stairs. & 0.0800 & 0.2321 & 0.1521 \\
\midrule

Fair & Many guards are standing in front of the starting line of a race. & Two men in business dress are standing by the side of a road. & 0.0800 & 0.2655 & 0.1855 \\
\midrule

Poor & 19 hurt in New Orleans shooting & Police: 19 hurt in NOLA Mother's Day shooting & 0.8800 & 0.6648 & 0.2152 \\
\midrule

Poor & I have a standing/sitting desk at work and really like it. & As mentioned by other responders, it turns out that using a standing desk isn't necessarily a perfect solution. & 0.4400 & 0.6800 & 0.2400 \\
\midrule

Poor & A skateboarder jumps off the stairs. & A dog jumps off the stairs. & 0.1600 & 0.5656 & 0.4056 \\

\end{longtable}
\normalsize

\vspace{-0.5cm}
\noindent The table provides three examples for each quartile of similarity score error made by the similarity model on the GLUE STS-B, grouped by the error range categories found in Table 11.

\textbf{Table 11: Performance Summary by Error Quartile}

\small
\begin{longtable}{p{3cm}p{4cm}}
\toprule
Category & Error Range \\
\midrule
\endfirsthead

\multicolumn{2}{c}{\textit{Table continued from previous page}} \\
\toprule 
Category & Error Range \\
\midrule
\endhead

\midrule
\multicolumn{2}{r}{{Continued on next page}} \\
\endfoot

\bottomrule
\endlastfoot

Excellent & 0.0000 - 0.0494 \\
\midrule

Good & 0.0497 - 0.1119 \\
\midrule

Fair & 0.1119 - 0.2036 \\
\midrule

Poor & 0.2042 - 0.6103 \\

\end{longtable}
\normalsize

\vspace{-0.5cm}
\noindent The table defines the error ranges per quartile for the model on the GLUE STS-B as ``Excellent'', ``Good'', ``Fair'', or ``Poor''.

\subsection{GLUE STS-B Human Similarity Score Guide}

\textbf{Table 12: Score Guide Provided to Human Scorers on the GLUE STS-B (Cer et al., 2017)}

\small
\begin{longtable}{p{3.5cm}p{9.5cm}}
\toprule
Score & Description/Example \\
\midrule
\endfirsthead

\multicolumn{2}{c}{\textit{Table continued from previous page}} \\
\toprule 
Score & Description/Example \\
\midrule
\endhead

\midrule
\multicolumn{2}{r}{{Continued on next page}} \\
\endfoot

\bottomrule
\endlastfoot

5 (Normalized: 1.0) & The two sentences are completely equivalent, as they mean the same thing. \\
\midrule

5: Example & The bird is bathing in the sink.

Birdie is washing itself in the water basin. \\
\midrule

4 (Normalized: 0.8) & The two sentences are mostly equivalent, but some unimportant details differ. \\
\midrule

4: Example & Two boys on a couch are playing video games.

Two boys are playing a video game. \\
\midrule

3 (Normalized: 0.6) & The two sentences are roughly equivalent, but some important information differs/missing. \\
\midrule

3: Example & John said he is considered a witness but not a suspect.

``He is not a suspect anymore.'' John said. \\
\midrule

2 (Normalized: 0.4) & The two sentences are not equivalent, but share some details. \\
\midrule

2: Example & They flew out of the nest in groups.

They flew into the nest together. \\
\midrule

1 (Normalized: 0.2) & The two sentences are not equivalent, but are on the same topic. \\
\midrule

1: Example & The woman is playing the violin.

The young lady enjoys listening to the guitar. \\
\midrule

0 (Normalized: 0.0) & The two sentences are completely dissimilar. \\
\midrule

0: Example & The black dog is running through the snow.

A race car driver is driving his car through the mud. \\

\end{longtable}
\normalsize

\vspace{-0.5cm}
\noindent The table is recreated from the paper introducing the Semantic Textual Similarity Benchmark by Cer et al. 2017, providing instructions for human scorers to rate the similarity of a sentence pair.

\subsection{Case Study 2: Generated Narratives by Outlet}

\textbf{Table 13: Narratives Generated by the Narrative Synthesis Tool in Case Study 2}

\footnotesize
\begin{longtable}{p{2.5cm}p{4cm}p{4cm}p{4cm}}
\toprule
Outlet & Narrative 1 & Narrative 2 & Narrative 3 \\
\midrule
\endfirsthead

\multicolumn{4}{c}{\textit{Table continued from previous page}} \\
\toprule 
Outlet & Narrative 1 & Narrative 2 & Narrative 3 \\
\midrule
\endhead

\midrule
\multicolumn{4}{r}{{Continued on next page}} \\
\endfoot

\bottomrule
\endlastfoot

Fox News & The dominant narrative is about the Montana Supreme Court blocking a ban on transgender individuals from using public restrooms that match their gender identity. The second dominant narrative is about the House Pentagon funding bill that would ban transgender individuals from serving in the military. & The dominant narrative is about the controversy surrounding transgender athletes particularly Lia Thomas, and the impact on female athletes. The second dominant narrative is about the political stance on transgender athletes, with many conservatives and Republicans speaking out against their inclusion in women's sports. & The dominant narrative is about the ongoing debate surrounding transgender rights, particularly regarding access to women's spaces and sports. This includes discussions about transgender women being `every bit as biological women' and the pushback against this perspective. The second dominant narrative is about the political polarization surrounding transgender issues, with conservatives and liberals having differing views on transgender rights. This includes discussions about transgender women being banned from Capitol Hill and the backlash against this decision. \\
\midrule

The Gateway Pundit & Transgender individuals are violent and dangerous. Transgender individuals are being discriminated against. & The struggle for transgender rights and acceptance in sports. The debate surrounding transgender individuals and their impact on women's sports. & Support for LGBTQ+ rights and protections. Concerns about LGTBQ+ indoctrination in schools and society. \\
\midrule

The Guardian & Criticism of the Census Bureau's handling of questions related to trans and gender diverse people. Praise for the Census Bureau's inclusion of questions related to trans and gender diverse people. & Support for student psychotherapist who won apology for being expelled from university for expressing unpopular views on gender identity. Criticism of the concept of `sigma males' and concerns about the impact of gender identity teaching on children in schools. & Support for trans rights and condemnation of transphobia following the murder of a trans model in Georgia. Criticism of Conservative Party's policies towards trans people, including claims of allowing bars on trans women and linking trans inclusion to girls' urinary tract infections. \\
\midrule

The New York Times & Discussion on the Supreme Court's decision on abortion rights. Focus on transgender rights and experiences. & The 4B movement that started in South Korea and its global impact. The Pope's visit to South Korea and its significance. & Athletic news coverage. Transgender athletes. \\

\end{longtable}
\normalsize

\vspace{-0.5cm}
\noindent The table types the narratives generated and displayed in Figure 7 for each news outlet analyzed.

\end{document}